\documentclass[conference]{IEEEtran}
\usepackage[cmex10]{amsmath}
\usepackage[left=0.75in,right=0.75in,bottom=0.75in,top=0.75in]{geometry}
\usepackage{amssymb}
\usepackage{epsfig,epsf,pstricks,pgf}
\usepackage{cite}

\usepackage{enumerate}

\usepackage{amsthm}
\usepackage{psfrag}

\usepackage{times}
\usepackage{epsfig}
\usepackage{amsmath}
\usepackage{amsfonts}
\usepackage{algorithm}
\usepackage{algorithmic}
\usepackage{graphicx}
\usepackage{amssymb}
\usepackage{amstext}
\usepackage{latexsym}
\usepackage{color}
\usepackage{ifthen}
\usepackage{multirow}
\usepackage{verbatim}
\usepackage{array,tabularx}
\usepackage{epsf,pstricks,pgf,pst-node}
\usepackage{color}
\usepackage{url}
\usepackage{subfigure}

\DeclareMathOperator*{\argmin}{\arg\!\min}

\newcommand{\mbf}{\mathbf}
\newcommand{\mcl}{\mathcal}

\newtheorem{theorem}{Theorem}
\newtheorem{lemma}{Lemma}

\theoremstyle{definition}
\newtheorem{definition}{Definition}
\newtheorem{example}{Example}
\newtheorem{remark}{Remark}

\IEEEoverridecommandlockouts
\begin{document}
\newgeometry{left=0.75in,right=0.75in,bottom=0.75in,top=1in}
%\newgeometry{margin=1in}

%\title{Separable Convex Cost Data Exchange Problem}
%\title{Efficient Algorithms for the Separable Convex Cost Data Exchange Problem}
%\title{Data Exchange Problems: Algorithms and Complexity}
\title{Efficient Algorithms for the Data Exchange Problem}
%\title{Data Exchange Problem with Separable Convex Cost}
\author{% aefraer \IEEEauthorrefmark{baer}
\IEEEauthorblockN{Nebojsa Milosavljevic, Sameer Pawar, Salim El Rouayheb, Michael Gastpar and Kannan Ramchandran}
\thanks{The material in this
paper appears in part in~\cite{milosavljevic2011deterministic,milosavljevic2011optimal,nebojsa_allerton}.}
\thanks{N. Milosavljevic, S. Pawar, M. Gastpar and K. Ramchandran are with the Department of Electrical Engineering and Computer Science, University
        of California, Berkeley, Berkeley, CA 94720 USA (e-mail:\{nebojsa,spawar, gastpar, kannanr\}@eecs.berkeley.edu).}
\thanks{S. El Rouayheb is with the Department of Electrical and Computer Engineering, Illinois Institute of Technology, Chicago, IL 60616 USA (e-mail: salim@iit.edu).}
\thanks{M. Gastpar is also with the School of Computer and Communication Sciences, EPFL, Lausanne, Switzerland (e-mail: michael.gastpar@epfl.ch).}
\thanks{This research was funded by the NSF grants
(CCF-0964018, CCF-0830788), a DTRA grant (HDTRA1-09-1-0032), and in part by an
AFOSR grants (FA9550-09-1-0120, FA9550-10-1-0567).}
}

% make the title area
\maketitle

%\linespread{0.97}

\begin{abstract}
In this paper we study the data exchange problem where a set of users is
interested in gaining access to a common file, but where each has only partial knowledge about it as side-information.
Assuming that the file is broken into packets, the side-information considered is in the form of
linear combinations of the file packets.
Given that the collective information of all the users is sufficient to allow
recovery of the entire file, the goal is for each user to gain access to the file while minimizing some communication cost.
We assume that users can communicate over a noiseless broadcast channel, and that the communication cost is a sum of each user's cost function
over the number of bits it transmits. For instance, the communication cost could simply be the total number of bits that needs to be transmitted.
In the most general case studied in this paper, each user can have any arbitrary convex cost function.
%A special case of this objective is the uniform cost, which essentially leads to a ``fair'' communication load distribution among the users.
We provide deterministic, polynomial-time algorithms (in the number of users and packets) which find an optimal communication scheme that minimizes the communication cost.
To further lower the complexity, we also propose a simple randomized algorithm inspired by our deterministic algorithm which is based on a random linear network coding
scheme.
\end{abstract} 
\section{Introduction}\label{sec:intro}
In recent years cellular systems  have witnessed significant improvements in terms of data rates, and are nearly
approaching the theoretical limits in terms of the physical layer spectral efficiency. At the same time, the rapid growth in
the popularity of data-enabled mobile devices, such as smart phones and tablets,
and the resulting explosion in demand for more throughput are challenging our abilities to deliver data,
even with the current highly efficient cellular systems. One of the major bottlenecks in scaling the throughput with the
increasing number of mobile devices is the ``last mile'' wireless link between the base station and the mobile devices -- a
resource that is shared among many users served within the cell. This motivates the study of paradigms where cell phone
devices can  cooperate among themselves to get the desired data in a peer-to-peer fashion without solely relying on the base station.

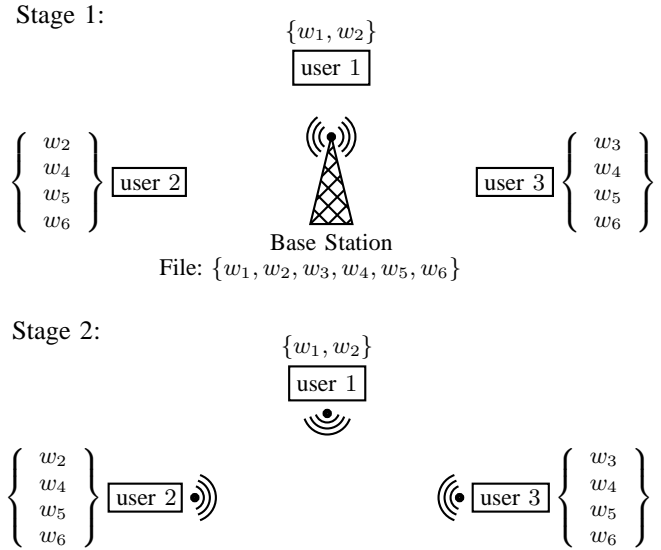
\begin{figure}
\begin{center}
\psset{unit=0.40mm}
\begin{pspicture}(0,-102)(170,85)
\rput(-5,75){Stage 1:}
\small{
\pspolygon[fillstyle=crosshatch*](78,6)(92,6)(85,35)
%\psdots(85,35)
\pscircle[linewidth=2pt](85,35){0.2}
\psarc{-}(85,35){8}{-55}{55}
\psarc{-}(85,35){6}{-55}{55}
\psarc{-}(85,35){4}{-55}{55}

\psarc{-}(85,35){8}{125}{235}
\psarc{-}(85,35){6}{125}{235}
\psarc{-}(85,35){4}{125}{235}

\rput(85,0){\small{Base Station}}
\rput(78,-9){File:~$\{w_1,w_2,w_3,w_4,w_5,w_6\}$}
\rput(-6,20){$\left\{
                \begin{array}{c}
                  w_2 \\
                  w_4 \\
                  w_5 \\
                  w_6 \\
                \end{array}
              \right\}$}
\psframe(12,15)(37,25)
\rput(25,20){user $2$}
\psframe(133,15)(159,25)
\rput(146,20){user $3$}
\rput(177,20){$\left\{
                 \begin{array}{c}
                   w_3 \\
                   w_4 \\
                   w_5 \\
                   w_6 \\
                 \end{array}
               \right\}
$}
\rput(85,70){$\left\{w_1,w_2\right\}
$}
\psframe(72,52)(98,64)
\rput(85,58){user $1$}
}
\rput(0,-105){
\normalsize{\rput(-5,75){Stage 2:}}
\small{
\rput(-6,20){$\left\{
                \begin{array}{c}
                  w_2 \\
                  w_4 \\
                  w_5 \\
                  w_6 \\
                \end{array}
              \right\}$}
\psframe(12,15)(37,25)
\rput(25,20){user $2$}
\psframe(133,15)(159,25)
\rput(146,20){user $3$}
\rput(177,20){$\left\{
                 \begin{array}{c}
                   w_3 \\
                   w_4 \\
                   w_5 \\
                   w_6 \\
                 \end{array}
               \right\}
$}
\rput(85,70){$\left\{w_1,w_2
              \right\}
$}
\psframe(72,52)(98,64)
\rput(85,58){user $1$}
\psarc{-}(40,20){8}{-70}{70}
\psarc{-}(40,20){6}{-70}{70}
\psarc{-}(40,20){4}{-70}{70}
%\psdots(41,20)
\pscircle[linewidth=2pt](41,20){0.2}

\psarc{-}(130,20){8}{110}{250}
\psarc{-}(130,20){6}{110}{250}
\psarc{-}(130,20){4}{110}{250}
%\psdots(129,20)
\pscircle[linewidth=2pt](129,20){0.2}

\psarc{-}(85,49){8}{200}{340}
\psarc{-}(85,49){6}{200}{340}
\psarc{-}(85,49){4}{200}{340}
%\psdots(85,48)
\pscircle[linewidth=2pt](85,48){0.2}

%\psline{->}(85,30)(85,52)

}
}
\end{pspicture}
\end{center}
\caption{An example of the data exchange problem. A base station has a file formed of six packets $w_1,\dots, w_6\in
\mathbb{F}_q$ and wants to  deliver it to three users over an unreliable wireless channel.
The base station
stops transmitting once all users collectively have all the packets, even if individually they have only subsets of the
packets (Stage 1). Users can then cooperate among themselves to recover their missing packets by broadcasting over a noiseless public channel (Stage 2).
It can be shown that the minimum number of symbols in $\mathbb{F}_q$ needed for the file recovery at all users is $5$.
A communication scheme that achieves this minimum is: user~$1$ transmits $w_1$, user~$2$ transmits $w_2+w_4$, while user~$3$ transmits $w_3$, $w_5$, $w_6$.
Now, if the goal is to allocate these $5$ transmissions to the users as uniformly as possible,
user~$1$ transmits $w_1$, user~$2$ transmits $w_2+w_4$, $w_5$, and user~$3$ transmits $w_3$, $w_6$.}
\label{fig:model_raw}
\end{figure} 

An example of such a setting is shown in Figure~\ref{fig:model_raw}, where a base station
wants to deliver the same file to multiple  geographically-close users over an unreliable wireless downlink.
In the example of Figure~\ref{fig:model_raw}, we assume that the file
consists of six equally sized packets $w_1$, $w_2$, $w_3$, $w_4$, $w_5$ and $w_6$ belonging to some finite field $\mathbb{F}_q$.
%We also assume throughout this work that the file packets are indivisible, \emph{i.e.}, it is not possible to split each packet
%into chunks of a smaller size.
Suppose that after a few initial transmission attempts by the base station, the three users individually receive only
parts of the file (see Figure~\ref{fig:model_raw}), but collectively have the entire file. Now, if all users
are in close vicinity  and can communicate with each other, then, it is much more desirable and efficient, in terms
of resource usage, to reconcile the file among users by letting all of them ``talk'' to each other without involving the  base station.
The cooperation among the users has the following advantages:
\begin{itemize}
\item Local communication among users has a smaller footprint in terms of interference, thus allowing one to use the shared resources (code, time or frequency)
freely without penalizing the base station's resources, \emph{i.e.}, higher resource reuse factor.
\item Transmissions within the close group of users is much more reliable than from the base station to any terminal due to geographical proximity of terminals.
\item This cooperation allows file recovery even when the connection to the base station is either unavailable after the initial phase of transmission, or it is too weak to meet the delay requirement.
\end{itemize}

Let us consider the example in Figure~\ref{fig:model_raw}, and let user~$1$, user~$2$ and user~$3$ transmit $R_1,R_2$ and $R_3$ symbols in $\mathbb{F}_q$, respectively. It can be shown that the minimum total number of symbols in $\mathbb{F}_q$ needed to recover the file is $5$. One possible communication scheme that achieves it is: user~$1$ transmits $w_1$, user~$2$ transmits $w_2+w_4$, while user~$3$ transmits $w_3$, $w_5$, $w_6$.
Note that the load of the communication of the system is unevenly distributed among the users,
\emph{i.e.}, user~$3$ transmits $3$ out of $5$ symbols in $\mathbb{F}_q$.
The next question we ask here is out of all communication schemes that deliver the entire file to the users in the minimum number of transmissions, which one distributes the load of communication to the users as fair as possible.
%This scenario backs the question of existence of a communication scheme that distributes the load of communication  as fair as %possible.
%The question we can ask here is whether there exists a scheme that distributes the load of communication  as fair as possible.
%Therefore, in this case, user~$3$ requires more battery consumption compared to the
%other users in order to complete its communication task. From that perspective, it is reasonable to construct
%a communication scheme that distributes communication load as fair as possible.
For instance, for the same minimum number
of transmissions, we can have the following scheme: user~$1$ transmits $w_1$, user~$2$ transmits $w_2+w_4$, $w_5$, and
user~$3$ transmits $w_3$, $w_6$. Intuitively, this scheme is more fair\footnote{To be precise, the fairness cost that we consider belongs to the broader class of separable convex costs that is studied in this work.}
%In Section~\ref{sec:gairness} we show that for the fixed number of transmissions, the cost $\sum_{i} R_i \log R_i$, where $i$ goes %over all the users, distributes communication load as evenly as possible among the users.}
than the previous one since it spreads the transmissions more uniformly among the users. And, it can be shown that such scheme minimizes a convex fairness cost.
%Since $R_i \log R_i$ is a convex function, the entire cost falls under the category of separable convex costs.

%It can be shown that such a scheme minimizes uniformity objective $\sum_{i=1}^3 R_i \log R_i$
%for a fixed number of transmissions equal to $5$ symbols in $\mathbb{F}_q$. In general, we can consider any separable convex
%cost function $\sum_{i=1}^3 \varphi_i (R_i)$, where $\varphi_i$ is a convex function.

In the example from Figure~\ref{fig:model_raw}, we considered only a simple form of side-information, where different users observe subset of uncoded ``raw'' packets  of the original file. Content distribution networks \cite{byers1999accessing, byers2002informed, liu2010uusee} are increasingly using codes, such as linear network codes or Fountain codes~\cite{luby2002lt}, to improve the system efficiency. In such  scenarios, the side-information representing
the partial knowledge gained by the users would be coded and in the form of linear combinations of the original file packets, rather than the raw packets themselves.
We refer to this model of side-information as a \emph{linear packet model}.
%The previous two cases of side-information (``raw'' and
%coded) can be regarded as special cases of the more general problem where the side-information has arbitrary
%correlation among the data observed by different users.

\subsection*{Contributions}

In this paper, we study the data exchange problem under the linear packet model and the separable convex communication cost.
Such cost captures all the communication objectives discussed earlier: 1. Minimization of the (weighted) sum of bits users need to exchange, 2. Fairness. Our contributions can be summarized as follows:
\begin{enumerate}
\item We propose a deterministic polynomial time algorithm for finding an optimal communication scheme w.r.t. the communication cost. An important step of this algorithm is to iteratively determine how much should each user transmit in an optimal scheme. We provide two methods to solve this problem. The first one is based on
    minimizing a submodular function, in which case the total complexity of the algorithm is $\mcl{O}((m^6\cdot N^3+m^7) \cdot \log N)$, where $m$ is the total number of users, and $N$ is the number of packets in the file. The second technique is based on subgradient methods, in which case the total complexity of the algorithm can be bounded by $\mcl{O}((N^2\cdot m^4\log m + N^5\cdot m^4) \cdot \log N)$ given that we use constant step size in the subgradient algorithm.
%    The first algorithm is fully combinatorial, and it's complexity is $\mcl{O}((m^6\cdot N^3+m^7) \cdot \log N)$, where $m$ is the total number of users, and $N$ is the number of packets in the file. The second algorithm uses subgradient methods to compute some certain steps of the first algorithm, and its overall complexity is
%    $\mcl{O}((N^2\cdot m^3\log m + N^5\cdot m^3) \cdot \log N)$.
\item We devise a randomized algorithm inspired by the deterministic scheme that reduces complexity to \mbox{$\mcl{O}(m \cdot N^4 \log N)$}. The randomized algorithm is based on a random linear network coding scheme, and it achieves the optimal
    number of transmissions with high probability. To be more precise, the probability of not achieving the optimum is inversely proportional to the underlying field size $|\mathbb{F}_q|$.
    %Therefore, depending upon the application and
    %whether or not we allow to be occasionally suboptimal, we can choose either deterministic or randomized algorithm.
    Our randomized algorithm can be regarded as a generalization of the algorithm proposed in~\cite{ozgul2011algorithm}, where the authors considered linear communication cost.
\item For the data exchange problem with additional capacity constraints on each user, we provide both deterministic and randomized algorithm of the same complexity as in 1. and 2.
\end{enumerate}
The challenging part of the deterministic algorithm is that the underlying optimization problem has exponential number of constraints coming from the cut-set bound region. By using combinatorial optimization techniques such as \emph{Dilworth truncation} and \emph{Edmonds' algorithm}, we devise an efficient, polynomial time solution.

\subsection*{Literature Overview}

The problem of reconciling a file among
multiple  wireless users having parts of it while minimizing  the  cost in terms of the total number of bits exchanged
is known in the literature as the {\em data exchange problem} and was introduced by El Rouayheb {\em et al.} in
\cite{SSS10}. A closely related problem was also studied by Csisz\'ar and Narayan in~\cite{CN04} where all users want to agree on a secret key in the presence of an eavesdropper who observes the entire communication.
A randomized algorithm for the data exchange problem was proposed in \cite{SSBE10},
while Tajbakhsh \emph{et al.} \cite{tajbakhsh2011generalized}
formulated this problem as a linear program (LP). The solution proposed in~\cite{tajbakhsh2011generalized}
is approximate.

The linear cost data exchange problem was studied by Ozgul {\em et al.} \cite{ozgul2011algorithm}, where the authors proposed a randomized algorithm. A deterministic polynomial time algorithm was proposed by Courtade and Wesel in~\cite{couratadealerton} concurrently to the authors' work~\cite{milosavljevic2011optimal}.
Minimum linear communication cost problem was also studied in the network coding literature.
Lun \emph{et al.}~\cite{lun2006minimum} proposed a polynomial time algorithm for the single source multicast problem
over a directed acyclic graph.

In~\cite{CXW10,gonen2012coded}, the authors considered a different version of the data exchange problem where users can only broadcast messages to their immediate neighbors.
In \cite{CXW10} it was shown that the problem is NP-hard, while an approximate solution is provided in~\cite{gonen2012coded}. In~\cite{lucani2009network}, Lucani \emph{et al.} considered the problem of data exchange when the channel between different users can have erasures.

%The literature so far considered only uncoded ``raw'' packet model.
%For the general correlation source model, in~\cite{milosavljevic2011optimal}, \cite{milosavljevic2012data} we proposed a deterministic polynomial-time
%algorithm that computes optimal communication rates of each user w.r.t. the linear cost.

In~\cite{CN04}, the authors posed a related security
problem referred to as  the  ``multi-terminal key agreement'' problem. They showed  that obtaining the file among the  users
in minimum number of bits exchanged over the public channel is sufficient to maximize the size of the secret key shared between the users. This result establishes  a connection between the Multi-party key agreement and the data exchange problem.

The rest of the paper is organized as follows. In Section~\ref{sec:model}, we describe the model and formulate the
optimization problem. In Section~\ref{sec:det_algorithm}, we provide a polynomial time algorithm that solves for
how many symbols in $\mathbb{F}_q$ should each user transmit. We start Section~\ref{sec:det_algorithm} by analyzing a linear
cost function, and then we extend our solution to any separable convex cost. In Section~\ref{sec:code}, we propose a polynomial time code construction.
In Section~\ref{sec:rand_alg}, we describe an algorithm based on random linear network coding approach,
that achieves the optimal communication cost. In Section~\ref{sec:cap_constraints}, we present a polynomial time solution to the
problem where each user additionally has capacity constraints, \emph{i.e.}, user~$i$ is not allowed to transmit more than $c_i$
symbols in $\mathbb{F}_q$. We conclude our work in Section~\ref{sec:conclusion}.

%\input{test.tex}
%!TEX root = Salim_DatExc.tex
\section{System Model and Preliminaries}\label{sec:model}
In this paper, we consider a setup with $m$ users that are interested in gaining access to a file.
The file is broken into $N$ linearly independent packets $w_1,\ldots,w_N$ each belonging to a field $\mathbb{F}_q$,
where $q$ is a power of some prime number.
Each user $i \in \mcl{M}\triangleq \{1,2,\ldots,m\}$  observes some collection of the linear combinations of the file packets as shown below.
\begin{align}
\mbf{x}_i = \mbf{A}_i \mbf{w}, \ i \in \mcl{M}, \label{model:eq1}
\end{align}
where $\mathbf{A}_i \in \mathbb{F}_q^{\ell_i \times N}$ is a given matrix, and $\mbf{w}=\left[
           \begin{array}{cccc}
             w_1 & w_2 & \ldots & w_N
           \end{array}\right]^T$ is a vector of the file packets.
In the further text, we refer to~\eqref{model:eq1} as a linear packet model.

Let us denote by $\mbf{v}_i$, a transmission of user~$i\in \mcl{M}$. In~\cite{CN04} it was shown that 
in order for each user to recover the file, interaction among them is not needed. Hence, without loss of generality, we can assume that $\mbf{v}_i$ is a function of user $i$'s initial observation.  We define
%It can be represented as a general mapping from the
%set of observations of user~$i$ over $\mathbb{F}_q^{\ell_i}$ to an arbitrary space. We define
\begin{align}
R_i \triangleq | \mbf{v}_i |_{q}
\end{align}
to be the size of user $i$'s transmission represented in number of symbols in $\mathbb{F}_q$.
To decode the file, user~$i$ collects transmissions of all the users and creates a decoding function
\begin{align}
\psi_i : \mathbb{F}_q^{\ell_i} \times \mathbb{F}_q^{R_1} \times \cdots \times \mathbb{F}_q^{R_m} \rightarrow \mathbb{F}_q^N,
\end{align}
that reconstructs the file, \emph{i.e.},
\begin{align}
\psi_i(\mbf{x}_i,\mbf{v}_1,\ldots,\mbf{v}_m)=\mbf{w}. \label{eq:decode}
\end{align}

\begin{definition}
A rate vector $\mbf{R}=(R_1,R_2,\ldots,R_m)$ is an {\em achievable data exchange (DE) rate vector} if there exists a communication scheme with transmitted messages $(\mbf{v}_1,\mbf{v}_2,\ldots,\mbf{v}_m)$ that satisfies \eqref{eq:decode} for all $i=1,\ldots,m$.
\end{definition}

\begin{remark}
Using cut-set bounds, it follows that all the achievable \emph{DE}-rate vectors necessarily belong to the following region
\begin{align}
\mcl{R}\triangleq \left\{\mbf{R} \in \mathbb{R}^m: R(\mcl{S})\geq N-\text{rank}(\mbf{A}_{\mcl{M} \setminus \mcl{S}}),~\forall \mcl{S}\subset \mcl{M} \right\}, \label{cut_set}
\end{align}
where
\begin{align}
R(\mcl{S}) \triangleq \sum_{i \in \mcl{S}} R_i,~~\text{and}~~
\mbf{A}_{\mcl{M} \setminus \mcl{S}} \triangleq \bigcup_{i \in \mcl{M} \setminus \mcl{S}} \mbf{A}_i.\nonumber
\end{align}
\end{remark}

\begin{theorem}\label{thm:netcode}
For a sufficiently large field size $|\mathbb{F}_q|$, any integer \emph{DE}-rate vector $\mbf{R} \in \mathbb{Z}^m$ that belongs to the cut-set region $\mcl{R}$,
can be achieved via linear network coding, \emph{i.e.},
it is sufficient for each user $i \in \mcl{M}$ to transmit $R_i$ properly chosen linear combinations of the data packets it observes.
\end{theorem}
Proof of Theorem~\ref{thm:netcode} is provided in Appendix~\ref{app:thm:netcode}.
In Section~\ref{sec:code} we show that any field size $|\mathbb{F}_q|$ larger than the number of users is sufficient to guarantee the existence of such solution. In general, finding the minimum field size can be a hard problem.

%The result here can be easily extended to rational rates by splitting the packets into equal chunks, and solving the
%problem using the integer solution.
%In Section~\ref{sec:optimal_split}, for minimizing the total number of transmissions over $\mathbb{F}_q$, we propose a polynomial time solution that finds the optimal packet split.
%To our knowledge, this communication cost is the only case when an efficient

%Finding the minimum field size $|\mathbb{F}_q|$ that allows linear network coding solution is a combinatorial problem, and in general it can be found
%by exhaustive search. However, in Section~\ref{sec:code} we point out that any field size larger than the number of users is sufficient to guarantee the existence of such solution.

%In order to achieve other non-integer \emph{DE}-rate vectors in $\mcl{R}$, it might be necessary to split packets over %$\mathbb{F}_q$ into smaller chunks. This is briefly examined in Section~\ref{sec:optimal_split}.

%However, in this work we assume that a
%packet is the smallest unit of information, and thus indivisible. For that reason we consider integer \emph{DE}-rate vectors that %belong to the region $\mcl{R}$.

In order for each user to recover the entire file, it is necessary to receive a sufficient number of linear combinations
of the other users' observations. Hence, $\mbf{v}_i$, $i \in \mcl{M}$, defined above is a vector of $R_i$ symbols in $\mathbb{F}_q$.
Therefore, $\mbf{v}_i$ can be written as follows
\begin{align}
\mbf{v}_i = \mbf{B}_i \mbf{x}_i = \mbf{B}_i  \mbf{A}_i  \mbf{w} = \mbf{U}_i  \mbf{w},
\end{align}
where $\mbf{B}_i$ is an $R_i \times \ell_i$ transmission matrix with elements belonging to $\mathbb{F}_q$.
In order for each user to recover the file, the transmission matrices $\mbf{B}_i$, $i \in \mcl{M}$ should satisfy,
\begin{align}
\text{rank}\left( \left[\begin{array}{c}
                             \mbf{A}_i \\
                             \mbf{U} \\
                           \end{array}
\right] \right) = N,~~~~\forall i \in \mcl{M}, \label{eq:decodemat}
\end{align}
where $\mbf{U}\triangleq \bigcup_{i=1}^m \mbf{U}_i$. Hence, the decoding function $\psi_i$ of user~$i \in \mcl{M}$
involves inverting the matrix given in~\eqref{eq:decodemat} in order to obtain $\mbf{w}$.

%This problem can be thought of as an extension of a multi-terminal Slepian-Wolf problem, where instead of
%having one decoder, all the users are interested in recovering the joint source. In such general case, each user~$i$
%observes $n$ memoryless copies of $\mathbf{x}_i$, where $\mathbf{x}_i$ is drawn from the joint distribution
%$P_{\mbf{x}_1,\ldots,\mbf{x}_m}$. Then, any achievable \emph{DE}-rate vector $\mbf{R} \in \mathbb{R}^m$ must satisfy
%cut-set constraints as in~\eqref{cut_set}, where the rank function is replaced with the entropy function.
%In the further text, we refer to this source model as the discrete memoryless multiple source (DMMS) model.
%We note that all the results we derived in this paper can be as well applied to the general source model.

In this work, we design a polynomial complexity scheme that achieves the file exchange among all the users while simultaneously minimizing a convex separable cost function $\sum_{i=1}^m \varphi_i (R_i)$, where $\varphi_i$, $i\in \mcl{M}$ is a non-decreasing convex function. Such assumption on monotonicity of function $\varphi_i$ is consistent with the nature of the problem at hand;  sending more bits is always more expensive than sending fewer.
From~\eqref{cut_set} and the above mentioned cost function, the problem considered in this work can be formulated
as the following optimization problem:
\begin{align}
&\min_{\mbf{R} \in \mathbb{Z}^m} \sum_{i=1}^m \varphi_i (R_i), \label{problem1} \\
&~~~~~\text{s.t.}~~R(\mcl{S})\geq N-\text{rank}(\mbf{A}_{\mcl{M} \setminus \mcl{S}}),~\forall \mcl{S}\subset \mcl{M}. \nonumber
\end{align}
Optimization problem~\eqref{problem1} is a convex integer problem with $2^m-2$ constraints.
It was shown in~\cite{CZ10} that only n of these constraints are active but the challenge is how to determine which of them are.
%, and therefore the convex optimization techniques would provide the optimal solution, but possibly not in polynomial time w.r.t. %the number of users.
Solving the optimization problem~\eqref{problem1} answers the question of how many symbols in $\mathbb{F}_q$ each user has to
transmit in an optimal scheme. In this paper we provide a polynomial time algorithm that solves problem~\eqref{problem1}.
Once we obtain an optimal rate allocation, the actual
transmissions of each user can be solved in polynomial time by using the algebraic network coding framework~\cite{KM03}, \cite{H05}. This is explained in Section~\ref{sec:code}.

\section{Deterministic Algorithm}  \label{sec:det_algorithm}

Our goal is to solve problem~\eqref{problem1} efficiently. To do so, we will split it into two subproblems:
\begin{enumerate}
\item Given a total budget constraint $\beta$, \emph{i.e.}, $R(\mcl{M})=R_1+R_2+\cdots R_m = \beta$, determine whether
      $\beta$ is feasible or not. If $\beta$ is feasible, find the feasible rate split among the users that will
      achieve the total budget $\beta$ and minimize the cost $\sum_{i=1}^m \varphi_i(R_i)$.
\item Find $\beta$ that minimizes the objective function.
\end{enumerate}
The bottleneck here is how to solve Problem~1 efficiently. The optimal value of $\beta$ can then be found using binary search (see Algorithm~\ref{alg:bin_search_int}) since the objective function is w.r.t. $\beta$.
First, let us identify these two problems by rewriting problem~\eqref{problem1} as follows
\begin{align}
\min_{\beta \in \mathbb{Z}_{+}} h(\beta), \label{de_problem}
\end{align}
where
\begin{align}
&h(\beta) \triangleq \min_{\mbf{R}\in \mathbb{Z}^m} \sum_{i=1}^m \varphi_i (R_i), \label{de_problem0} \\
&~~\text{s.t.}~~R(\mcl{M})=\beta,~~R(\mcl{S})\geq N-\text{rank}(\mbf{A}_{\mcl{M} \setminus \mcl{S}}),
~\forall \mcl{S}\subset \mcl{M}. \nonumber
\end{align}
Note that the optimizations~\eqref{de_problem} and~\eqref{de_problem0} are associated with Problem~2 and Problem~1 defined above, respectively. Next we will explain our approach to solving these two problems.

\subsection{Optimization with a given sum-rate budget $\beta$}

Now, let us focus on the set of constraints of optimization problem~\eqref{de_problem0}. By substituting
$\mcl{S}$ with $\mcl{M}\setminus \mcl{S}$, we obtain
\begin{align}
R(\mcl{M}) &= \beta, \nonumber \\
R(\mcl{M} \setminus \mcl{S}) &= R(\mcl{M}) - R(\mcl{S}) = \beta - R(\mcl{S}) \nonumber \\
                             &\geq N-\text{rank}(\mbf{A}_{\mcl{S}}),~\forall \mcl{S} \subset \mcl{M},~\mcl{S} \neq \emptyset. \label{de_constraints}
\end{align}
Therefore, optimization problem~\eqref{de_problem0} can be equivalently represented as follows
\begin{align}
&h(\beta) = \min_{\mbf{R}\in \mathbb{Z}^m} \sum_{i=1}^m \varphi_i (R_i), \label{de_problem1} \\
&~~\text{s.t.}~~R(\mcl{M})=\beta, \nonumber \\
&~~~~~~~~R(\mcl{S})\leq \beta - N + \text{rank}(\mbf{A}_{\mcl{S}}),
~\forall \mcl{S}\subset \mcl{M},~\mcl{S} \neq \emptyset. \nonumber
\end{align}

Before we go any further, let us introduce some concepts from combinatorial optimization theory.
\begin{definition}[Polyhedron]
Let $f_{\beta}$ be a set function defined over set $\mcl{M}=\{1,2,\ldots,m\}$, \emph{i.e.}, $f_{\beta}:2^{\mcl{M}}\rightarrow \mathbb{Z}$, where $2^{\mcl{M}}$ is the power set of $\mcl{M}$.
Then the \emph{polyhedron} $P(f_{\beta})$ and the \emph{base polyhedron} $B(f_{\beta})$ of $f_{\beta}$ are defined as follows.
\begin{align}
P(f_{\beta}) & \triangleq \{\mbf{R} \in \mathbb{Z}^m~|~R(\mcl{S})\leq f_{\beta}(\mcl{S}),~\forall \mcl{S}\subseteq \mcl{M} \}, \label{f:poyh} \\
B(f_{\beta}) & \triangleq \{\mbf{R} \in P(f_{\beta})~|~R(\mcl{M})=f_{\beta}(\mcl{M})\} \label{co_f:base}.
\end{align}
\end{definition}
Note that the set of constraints of problem~\eqref{de_problem1}, for any fixed $\beta\in \mathbb{Z}_{+}$, constitutes the base polyhedron $B(f_{\beta})$ of the set function
\begin{align}
f_{\beta}(\mcl{S})=
\begin{cases}
\beta - N +\text{rank}(\mbf{A}_{\mcl{S}})  &  \text{if}~\mcl{S}\subset \mcl{M},~\mcl{S}\neq \emptyset \label{fcn:f_star} \\
\beta                           &  \text{if}~\mcl{S}=\mcl{M}, \\
0                               &  \text{if}~\mcl{S}=\emptyset.
\end{cases}
\end{align}

\begin{example}\label{example1}
Let us consider the source model from Figure~\ref{fig:model_raw}, where the three users observe the following parts of the file
\mbox{$\mbf{w} = \left[
             \begin{array}{cccccc}
               w_1 & w_2 & w_3 & w_4 & w_5 & w_6
             \end{array}
           \right]^T$:}
\begin{align}
\mbf{x}_1&=\left[
             \begin{array}{cc}
               w_1 & w_2
             \end{array}
           \right]^T, \nonumber \\
\mbf{x}_2&=\left[
             \begin{array}{cccc}
               w_2 & w_4 & w_5 & w_6
             \end{array}
           \right]^T, \nonumber \\
\mbf{x}_3&=\left[
             \begin{array}{cccc}
               w_3 & w_4 & w_5 & w_6
             \end{array}
           \right]^T. \label{example:source_model}
\end{align}
For $\beta=4$, the base polyhedron $P(f_4)$ is defined by the following set of inequalities:
\begin{align}
&R_1\leq f_4(\{1\})=0,~~R_2\leq f_4(\{2\})=2,~~R_3\leq f_4(\{3\})=2, \nonumber \\
&R_1+R_2\leq f_4(\{1,2\})=3,~~R_1+R_3\leq f_4(\{1,3\})=4,\nonumber \\
&R_2+R_3\leq f_4(\{2,3\})=3, \nonumber \\
&R_1+R_2+R_3 \leq f_4(\{1,2,3\})=4. \label{fcn:f_4}
\end{align}
It can be verified that no rate vector $(R_1,R_2,R_3) \in P(f_4)$ exists such that
$R_1+R_2+R_3=4$. Therefore, $B(f_4)=\emptyset$.
On the other hand, for $\beta=5$, the polyhedron $P(f_5)$ is defined as follows
\begin{align}
&R_1\leq f_5(\{1\})=1,~~R_2\leq f_5(\{2\})=3,~~R_3\leq f_5(\{3\})=3, \nonumber \\
&R_1+R_2\leq f_5(\{1,2\})=4,~~R_1+R_3\leq f_5(\{1,3\})=5, \nonumber \\
&R_2+R_3\leq f_5(\{2,3\})=4, \nonumber \\
&R_1+R_2+R_3\leq f_5(\{1,2,3\})=5. \label{fcn:f_5}
\end{align}
It can be easily verified that the rate vector $R_1=1$, $R_2=3$, $R_3=1$ belongs to the polyhedron $P(f_5)$.
Therefore, $B(f_5) \neq \emptyset$.
\end{example}

Summarizing the discussion so far, the optimization problem~\eqref{de_problem1} is equivalent to
\begin{align}
\min_{\mbf{R} \in \mathbb{Z}^m} \sum_{i=1}^m \varphi_i (R_i),~~\text{s.t.}~~\mbf{R}\in B(f_{\beta}), \label{de_problem2}
\end{align}
where $f_{\beta}$ is defined in~\eqref{fcn:f_star}.
For now, let us assume that parameter $\beta$ is chosen such that the optimization problem~\eqref{de_problem2} is feasible, \emph{i.e.}, $B(f_{\beta}) \neq \emptyset$. We will explain later how the condition $B(f_{\beta}) \neq \emptyset$ can be efficiently verified.

The main idea behind solving the optimization problem in~\eqref{de_problem2} efficiently, is to utilize the combinatorial properties of the set function $f_{\beta}$.
\begin{definition} \label{def:inter_submodular}
We say that a set function $f:2^{\mcl{M}}\rightarrow \mathbb{Z}$ is \emph{intersecting submodular} if
\begin{align}
&f(\mcl{S})+f(\mcl{T})\geq f(\mcl{S}\cup \mcl{T})+f(\mcl{S}\cap \mcl{T}), \nonumber \\
&~~~~~~\forall \mcl{S},\mcl{T} \subseteq \mcl{M}~~\text{s.t.}~~\mcl{S}\cap \mcl{T}\neq \emptyset. \label{eq:inter_sub}
\end{align}
When the inequality conditions in~\eqref{eq:inter_sub} are satisfied for all sets $\mcl{S}, \mcl{T} \subseteq \mcl{M}$,
the function $f$ is \emph{fully submodular}.
\end{definition}
\begin{lemma}\label{lm:f}
The function $f_{\beta}$ is \emph{intersecting submodular} for any $\beta$. When $\beta\geq N$, $f_{\beta}$ is fully submodular.
\end{lemma}
Proof of Lemma~\ref{lm:f} is provided in Appendix~\ref{app:lm:f}.
\begin{theorem}[Dilworth Truncation~\cite{F05}]\label{thm:dilw}
For every intersecting submodular function $f_{\beta}$ there exists a fully submodular
function $g_{\beta}$ such that both functions have the same polyhedron, i.e.,
\mbox{$P(g_{\beta})=P(f_{\beta})$}, and $g_{\beta}$ can be expressed as
\begin{align}
g_{\beta}(\mcl{S})=\min_{\mcl{P}}\left\{\sum_{\mcl{V}\in \mcl{P}}f_{\beta}(\mcl{V}) : \text{$\mcl{P}$ is a partition of $\mcl{S}$}\right\}.  \label{dilw1}
\end{align}
The function $g_{\beta}$ is called the \emph{Dilworth truncation} of $f_{\beta}$.
\end{theorem}
%We denote by $\mcl{P}^{*}_{\beta}$ an optimal partition of the set $\mcl{S} = \mcl{M}$ in~\eqref{dilw1} for any given $\beta$.
The base polyhedron of any fully submodular function always exists, \emph{i.e.}, there exists a rate vector $\mbf{R}$ such that $R(\mcl{M})=g_{\beta}(\mcl{M})$. Since, $P(g_{\beta})=P(f_{\beta})$, it follows that $B(g_{\beta})=B(f_{\beta})$ whenever $g_{\beta}(\mcl{M})=f_{\beta}(\mcl{M})=\beta$, \emph{i.e.}, when $B(f_{\beta})\neq \emptyset$ which implies feasibility of the optimization problem~\eqref{de_problem2}.

Continuing with Example~\ref{example1}, the Dilworth truncation of the set function $f_4$ is given by
\begin{align}
&g_4(\{1\})=0,~g_4(\{2\})=2,~g_4(\{3\})=2, \nonumber \\
&g_4(\{1,2\})=2,~g_4(\{1,3\})=2,~g_4(\{2,3\})=3, \nonumber \\
&g_4(\{1,2,3\})=3. \label{fcn:g_4}
\end{align}
Note that \mbox{$f_4(\{1,2,3\})\neq g_4(\{1,2,3\})$}, and hence, $\beta=4$ is not a feasible sum-rate for the problem~\eqref{de_problem2}.
On the other hand, for $\beta=5$, Dilworth truncation of a set function $f_5$ is given by
\begin{align}
&g_5(\{1\})=1,~g_5(\{2\})=3,~g_5(\{3\})=3, \nonumber \\
&g_5(\{1,2\})=4,~g_5(\{1,3\})=4,~g_5(\{2,3\})=4, \nonumber \\
&g_5(\{1,2,3\})=5. \label{fcn:g_5}
\end{align}
Now, \mbox{$f_5(\{1,2,3\} = g_5(\{1,2,3\})=\beta=5$} which indicates that $\beta=5$ is a feasible sum-rate for the problem~\eqref{de_problem2}.
Hence, the optimization problem~\eqref{de_problem2} can be written as
\begin{align}
\min_{\mbf{R} \in \mathbb{Z}^m} \sum_{i=1}^m \varphi_i(R_i),~~\text{s.t.},~~\mbf{R}\in B(g_{\beta}) \label{de_sum2}
\end{align}
provided that $g_{\beta}(\mcl{M})=\beta$.
\begin{remark}\label{rmk:feasibility}
Parameter $\beta$ is feasible w.r.t. the problem~\eqref{de_problem2} if $g_{\beta}(\mcl{M})=\beta$. Otherwise,
$g_{\beta}(\mcl{M})<\beta$. This is the direct consequence of the Dilworth truncation~\eqref{dilw1}.
\end{remark}
Depending upon the cost function $\sum_{i=1}^m \varphi(R_i)$, in the sequel, we provide several algorithms that can efficiently solve problem~\eqref{de_problem}.
First, we analyze a special case when the cost function is linear,
\begin{align}
\varphi_i(R_i)=\alpha_i R_i,~~\alpha_i>0,~~\forall i \in \mcl{M}. \label{ch1:linear_cost}
\end{align}
The condition $\alpha_i>0$, $i \in \mcl{M}$ ensures that $\varphi_i$ is a non-decreasing function.

\subsection{Linear Cost - Edmonds' Algorithm}

When the cost function is linear, the optimization problem~\eqref{de_sum2} has the following form
\begin{align}
\min_{\mbf{R} \in \mathbb{Z}^m} \sum_{i=1}^m \alpha_i R_i,~~\text{s.t.},~~\mbf{R}\in B(g_{\beta}). \label{de_sum3}
\end{align}
Due to the submodularity of function $g_{\beta}$, the optimization problem~\eqref{de_sum3} can be solved analytically
using Edmonds' greedy algorithm~\cite{E70} (see Algorithm~\ref{alg:edm}).

\begin{algorithm}
\caption{Edmonds' Algorithm}
\label{alg:edm}
\begin{algorithmic}[1]
\STATE Set $j(1),j(2),\ldots, j(m)$ to be an ordering of $\{1,2,\ldots,m\}$ such that \mbox{$\alpha_{j(1)}\leq \alpha_{j(2)}\leq \cdots \leq \alpha_{j(m)}$}
\STATE Initialize $\mbf{R}^{*}=\mbf{0}$.
\FOR {$i=1$ to $m$}
\STATE  ~\vspace{-0.22in}
        \begin{align}
        &R^{*}_{j(i)} = g_{\beta}(\{j(1),j(2),\ldots,j(i)\}) \nonumber \\\
        &~~~~~~~~~- g_{\beta}(\{j(1),j(2),\ldots,j(i-1)\}). \nonumber
        \end{align}
\ENDFOR
\end{algorithmic}
\end{algorithm}
The greediness of this algorithm is
reflected in the fact that each update of the rate vector is sum-rate optimal:
\begin{align}
R^{*}_{j(1)} &= g_{\beta}(\{j(1)\}) \nonumber\\
R^{*}_{j(1)}+R^{*}_{j(2)} &= g_{\beta}(\{j(1),j(2)\}) \nonumber\\
&~\vdots \nonumber \\
\sum_{i=1}^m R^{*}_{j(i)} &= g_{\beta}(\{j(1),\ldots,j(m)\}). \label{edmonds:greedyness}
\end{align}
In other words, at each iteration,
the individual user's rate update reaches the boundary of polyhedron $P(g_{\beta})$.
Optimality of this approach is the direct consequence of submodularity of function $g_{\beta}$~\cite{E70}.
\begin{remark}\label{rmk:base}
The optimal rate vector $\mbf{R}^{*}$ belongs to the base polyhedron $B(g_{\beta})$. In other words,
\begin{align}
\sum_{i=1}^m R^{*}_i = g_{\beta}(\mcl{M}).
\end{align}
\end{remark}

\begin{remark}
The complexity of Edmonds' algorithm is \mbox{$\mcl{O}(m \cdot \vartheta)$}, where $\vartheta$ is the complexity of
computing function $g_{\beta}(\mcl{S})$ for any given set $\mcl{S} \subseteq \mcl{M}$.
\end{remark}

\begin{figure}[h]
\begin{center}
\includegraphics[scale=0.45]{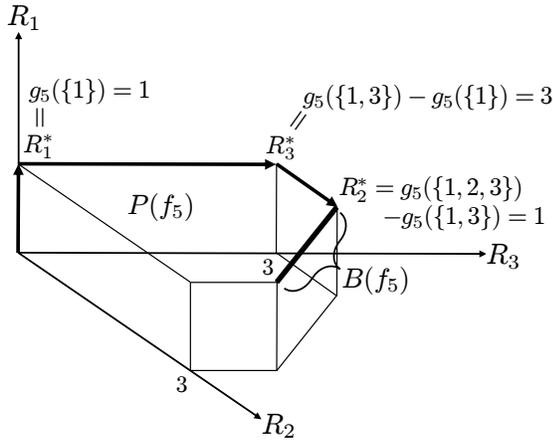}
\end{center}
\vspace{-0.2in}
\caption{Edmonds' algorithm applied to the three-user problem described in Example~\ref{example1}, with the
         cost function $R_1+3R_2+2R_3$. To minimize the cost, the order in which we greedily update
         communication rates should be increasing w.r.t. the weight vector, i.e., $1\rightarrow 3 \rightarrow 2$.
         The optimal \emph{DE}-rate vector is $R^{*}_1=1$, $R^{*}_2=1$, $R^{*}_3=3$.} \label{fig:example1}
\end{figure}

\begin{example}\label{example1}
Let us consider the same source model as in Example~\ref{example1}, and let the cost function be $R_1+3R_2+2R_3$, and $\beta=5$.
The intersecting submodular function $f_{\beta}$, and its Dilworth truncation $g_{\beta}$ are given in~\eqref{fcn:f_5} and~\eqref{fcn:g_5}, respectively.
The rate vector is updated in an increasing order w.r.t. the weight vector. In this case,
the order is $1\rightarrow 3 \rightarrow 2$ (see Figure~\ref{fig:example1}).
\end{example}

The main problem in executing Edmonds' algorithm efficiently is that the function $g_{\beta}$ is not available analytically.
To compute this function for any given set $\mcl{S} \subseteq \mcl{M}$ we need to solve minimization problem~\eqref{dilw1}.
Such minimization has to be performed over all partitions of the set $\mcl{S}$, which annuls the efficiency of the proposed method.

To overcome this problem note that we have access to the function $f_{\beta}$ (see~\eqref{fcn:f_star}), and by Theorem~\ref{thm:dilw}, we know that $P(g_{\beta}) = P(f_{\beta})$.
As pointed out before, each rate update reaches the boundary of polyhedron $P(g_{\beta})$ (see~\eqref{edmonds:greedyness}). Since we don't explicitly have function $g_{\beta}$, this polyhedron boundary can be calculated by applying the Dilworth truncation formula~\eqref{dilw1}. For the three-user problem in Example~\ref{example1} this procedure would go as follows:
\begin{align}
&R^{*}_1 = f_5(\{1\})=1, \nonumber \\
&R^{*}_3 = \min\{f_5(\{1,3\})-R^{*}_1,f_5(\{3\})\} = 3, \nonumber \\
&R^{*}_2 = \min\{f_5(\{1,2,3\})-R^{*}_1-R^{*}_3,f_5(\{1,2\})-R^{*}_1, \nonumber \\
&~~~~~~~~~~~~~f_5(\{2,3\})-R^{*}_3,f_5(\{2\})\} = 1. \nonumber
\end{align}

Generalization of this procedure to an arbitrary number of users is shown in Algorithm~\ref{alg:modedm}. We refer the interested reader to references~\cite{F05} and~\cite{NKI10} where this algorithm is explained in more details for an arbitrary intersecting submodular functions.

\begin{algorithm}
\caption{Modified Edmonds' Algorithm}
\label{alg:modedm}
\begin{algorithmic}[1]
\STATE Set $j(1),j(2),\ldots, j(m)$ to be an ordering of $\{1,2,\ldots,m\}$ such that \mbox{$\alpha_{j(1)}\leq \alpha_{j(2)}\leq \cdots \leq \alpha_{j(m)}$}
\STATE Initialize $\mbf{R}^{*}=\mbf{0}$.
\FOR {$i=1$ to $m$}
\STATE  ~\vspace{-0.22in}
        \begin{align}
        &R^{*}_{j(i)}=\min_{\mcl{S}} \{f_{\beta}(\mcl{S} \cup \{j(i)\})-R^{*}(\mcl{S}) :  \nonumber \\
        &~~~~~~~~~~~~~~~\mcl{S}\subseteq \{j(1),j(2),\ldots,j(i-1)\} \} \label{sub_mod}
        \end{align}
\ENDFOR
\end{algorithmic}
\end{algorithm}
In each iteration $i$, the minimization problem~\eqref{sub_mod} is over all subsets of $\{j(1),\ldots,j(i)\}$.
Using the fact that all the subsets considered in \eqref{sub_mod} contain a common element $j(i)$ it is easy to see that $f_{\beta}(\mcl{S})-R^{*}(\mcl{S})$ is fully submodular over the domain set $\{j(1),j(2),\ldots,j(i-1)\}$. Now the polynomial time solution of Algorithm~\ref{alg:modedm} follows from the fact that
minimization of a fully submodular function can be done in polynomial time~\cite{orlin2009faster}.

%
%follows from the fact that all the
%subsets of $\{j(1),j(2),\ldots,j(i)\}$ in iteration $i$ of Algorithm~\ref{alg:modedm} have a common element $j(i)$ (see~\eqref{sub_mod}), and thus, they all intersect.}
%In other words, the function $f_{\beta}$ over all the subsets $\mcl{S} \subseteq $, such that $j(i) \in \mcl{S}$, is fully submodular.
%It can be easily verified that the set function $R^{*}(\mcl{S})$ is modular. Since the difference of fully submodular and modular function results in a fully submoular function, it immediately follows that the function $f_{\beta}(\mcl{S})-R^{*}(\mcl{S})$ defined in~\eqref{sub_mod} is fully submodular.

\begin{remark}\label{rmk:complex}
The complexity of Algorithm~\ref{alg:modedm} is \mbox{$\mcl{O}(m\cdot SFM(m))$}, where $SFM(m)$ is the complexity of minimizing submodular function. The best known algorithm to our knowledge is proposed by Orlin in~\cite{orlin2009faster}, and has complexity $\mcl{O}(m^5\cdot \gamma+m^6)$, where $\gamma$ is complexity of computing the submodular function. For the submodular function defined in~\eqref{sub_mod}, $\gamma$ equals to the complexity of computing rank, and it is a function of the file size $N$. When users observe linear combinations of the file packets, the rank over $\mathbb{F}_q$ can be computed by Gaussian elimination in $\mcl{O}(N^3)$ time. For the ``raw'' packet model, rank computation reduces to counting distinct packets, and therefore its complexity is $\mcl{O}(N)$.
\end{remark}

\begin{remark}
From Remark~\ref{rmk:feasibility} and the fact that Edmonds' algorithm provides a rate vector with sum-rate $g_{\beta}(\mcl{M})$, it immediately follows that if Algorithm~\ref{alg:modedm} outputs
a rate vector $\mbf{R}^{*}$ such that $R^{*}(\mcl{M})<\beta$, then $B(f_{\beta}) = \emptyset$, and such $\beta$ is not a feasible
sum-rate w.r.t. the problem~\eqref{de_problem2}. Hence, for any given $\beta$, the feasibility of such sum-rate can be verified
in $\mcl{O}(m \cdot SFM(m))$ time.
\end{remark}

\subsection{Finding the optimal value of $\beta$}

So far we have shown how to compute function $h(\beta)$ defined in~\eqref{de_problem1} for any $\beta$ when $\varphi_i(R_i) = \alpha_i R_i$. To complete our solution, \emph{i.e.}, to solve the problem defined in~\eqref{de_problem}, it remains to show how to minimize function $h(\beta)$ efficiently.
\begin{theorem}\label{thm:convex}
Function $h(\beta)$, defined in~\eqref{de_problem0}, is convex when $\beta$ is a feasible sum-rate w.r.t. the optimization problem~\eqref{de_problem0}.
\end{theorem}
Proof of Theorem~\ref{thm:convex} is provided in Appendix~\ref{app:thm:convex}.
%Observe that Theorem~\ref{thm:convex} holds for a general separable convex cost $\sum_{i=1}^m \varphi_i(R_i)$, where
%$\varphi_i$, $i \in \mcl{M}$, is a non-decreasing convex function.

In order to minimize function $h$,
first, we identify the set of sum-rates $\beta$ that are feasible w.r.t. the
problem~\eqref{de_problem}. More precisely, we need to find the minimum sum-rate, since every $\beta$ that is larger than or equal to such value is feasible as well. Hence, we proceed by analyzing the sum-rate objective, \emph{i.e.}, when $\varphi_i(R_i) = R_i$.

For any fixed parameter $\beta \in \mathbb{Z}_{+}$, Algorithm~\ref{alg:modedm} provides an optimal rate allocation w.r.t. the linear cost.
It is only left to find $\beta$ that minimizes $h(\beta)$ in~\eqref{de_problem}.
Let us first consider the sum-rate cost, \emph{i.e.}, $\varphi_i(R_i)=R_i$.
From the equivalence of the Algorithms~\ref{alg:edm} and ~\ref{alg:modedm}, and from Remark~\ref{rmk:base} it follows that
for any given parameter $\beta$, the output rate vector $\mbf{R}^{*}$ of Algorithm~\ref{alg:modedm} satisfies
\begin{align}
\sum_{i=1}^m R_i^{*} = g_{\beta}(\mcl{M}).
\end{align}
Thus, for a randomly chosen parameter $\beta$ we can verify whether it is feasible w.r.t. the problem~\eqref{de_problem1}
by applying Remark~\ref{rmk:feasibility}, \emph{i.e.}, if $\sum_{i=1}^m R_i^{*}=\beta$, then such sum-rate can be achieved.
Therefore, we can apply a simple binary search algorithm to find the minimum sum-rate.
Note that the minimum sum-rate is always less than or equal to the file size $N$. Hence, we can confine our search accordingly (see Algorithm~\ref{alg:bin_search_int}).

\begin{algorithm}[h]
\caption{Minimum Sum-Rate Algorithm (binary search)}
\label{alg:bin_search_int}
\begin{algorithmic}[1]
\STATE Initialize $\beta_{start}=0$, $\beta_{end}=N$.
\WHILE {$\beta_{end}-\beta_{start}>1$}
\STATE $\beta = \lceil \frac{\beta_{start}+\beta_{end}}{2} \rceil$.
\STATE Execute Algorithm~\ref{alg:modedm} with parameter $\beta$.
\STATE \textbf{if}~~$\sum_{i=1}^m R^{*}_i = \beta$, \textbf{then}
\STATE ~~~$\beta_{end} = \beta$.
\STATE \textbf{else}~~$\beta_{start} = \beta$.
\ENDWHILE
\STATE $\beta_{end}$ is the minimum sum-rate.
\end{algorithmic}
\end{algorithm}

\begin{remark}
The complexity of Algorithm~\ref{alg:bin_search_int} is \mbox{$\mcl{O}(m \cdot SFM(m)\cdot \log N)$}.
\end{remark}

For the general linear cost function $\varphi_i(R_i) = \alpha_iR_i$, by Theorem~\ref{thm:convex}, $h(\beta)$ is convex for $\beta$ greater than
the minimum sum-rate (obtained from Algorithm~\ref{alg:bin_search_int}). In Section~\ref{sec:generalcost}, Lemma~\ref{lm:cost_cap}, we show that the search
space for $\beta$ that minimizes function $h$ can be limited to the file size $N$. Hence, in order to solve the minimization problem~\eqref{de_problem}
we can apply a simple binary search algorithm that finds the minimum of $h(\beta)$ by looking for a slope change in function $h$.

\begin{algorithm}[h]
\caption{Minimum Linear Cost Algorithm}
\label{alg:linear_cost}
\begin{algorithmic}[1]
\STATE Initialize $\beta_{start}=\beta_{sum}^{*}$, $\beta_{end}=N$, where $\beta_{sum}^{*}$ is the minimum sum-rate obtained from Algorithm~\ref{alg:bin_search_int}.
\STATE $\beta = \lceil \frac{\beta_{start}+\beta_{end}}{2} \rceil$.
\STATE Execute Algorithm~\ref{alg:modedm} for $\beta-1$, $\beta$, and $\beta+1$.
\STATE \textbf{if} $h(\beta)\leq h(\beta-1)$ and $h(\beta)\leq h(\beta+1)$, \textbf{then}
\STATE ~~~$\mbf{R}^{*}$ that corresponds to the sum-rate $\beta$ is an optimal rate allocation
\STATE \textbf{else if} $h(\beta-1)\geq h(\beta) \geq h(\beta+1)$, \textbf{then}
\STATE ~~~$\beta_{start} = \beta+1$.
\STATE \textbf{else} $\beta_{end}=\beta-1$.
\STATE Go to Step~2.
\end{algorithmic}
\end{algorithm}
\begin{remark}
Since for any fixed $\beta$, $h(\beta)$
can be found by using Algorithm~\ref{alg:modedm}, and $\beta_{sum}^{*}$ can be found by applying Algorithm~\ref{alg:bin_search_int},
the complexity of Algorithm~\ref{alg:linear_cost} is \mbox{$\mcl{O}(m \cdot SFM(m)\cdot \log N)$}.
\end{remark}

\subsection{Using Subgradient Methods to Solve Step 4 of Algorithm~\ref{alg:modedm}} \label{sec:subgradients}

In this section we propose an alternative solution to the minimization problem~\eqref{sub_mod} in Algorithm~\ref{alg:modedm}
that does not involve minimization of a submodular function.
%In this section we revisit Algorithm~\ref{alg:modedm} from Chapter~\ref{ch1}.
%As we pointed out in Remark~\ref{rmk:complex}, the complexity of this algorithm is $\mcl{O}(m^6\cdot N^3+m^7)$,
%and its most complex subroutine is minimizing of submodular function. The question we ask here is whether it is
%possible to run Algorithm~\ref{alg:modedm} without this minimization step. In Section~\ref{ch2:sec:noncomb} we applied
%non-combinatorial methods to the seemingly hard problems, and managed to obtain efficient solutions. Here, we consider a
%mixture of combinatorial and non-combinatorial techniques in order to answer the above question.
The underlying linear optimization problem has the following form
\begin{align}
\min_{\mbf{R}\in \mathbb{Z}^m} \sum_{i=1}^m\alpha_i R_i,~~\text{s.t.}~\mbf{R}\in B(f_{\beta}), \label{ch3:min_sum}
\end{align}
given that $\beta$ is a feasible sum-rate.
Without loss of generality, let us assume that $\alpha_1\leq \alpha_2\leq \cdots \leq \alpha_m$. In this case, the minimization
in Step~3 of Algorithm~\ref{alg:modedm} can be written as
\begin{align}
&R^{*}_i=\min_{\mcl{S}} \{f_{\beta}(\mcl{S})-R^{*}(\mcl{S}) : i\in \mcl{S}, \nonumber \\
&~~~~~~~~~~~~~~~\mcl{S}\subseteq \{1,2,\ldots,i\} \},~~i=1,2,\ldots,m. \label{ch3:sub_mod}
\end{align}
Minimization~\eqref{ch3:sub_mod} can be interpreted as a maximal update along the $i^{th}$ coordinate
such that $R^{*}_i$ still belongs to polyhedron $P(f_{\beta})$. This problem can be separately formulated as
the following minimization problem
\begin{align}
&R_i^{*} = \max_{\mbf{R} \in \mathbb{R}^i} R_i, \label{ch3:primal} \\
&~\text{s.t.}~R_k\geq R_k^{*},~k=1,2,\ldots,i-1, \nonumber \\
&~~~~~R(\mcl{S}\cup \{i\})\leq f_{\beta}(\mcl{S} \cup \{i\}),~~\forall \mcl{S} \subseteq \{1,2,\ldots,i-1\}. \nonumber
\end{align}
Note that in an optimal solution, the condition $R_k\geq R_k^{*}$, $k=1,\ldots,i-1$, holds with equality because any possible
increase of $R_k$ can lead to the smaller value of $R_i$. Moreover, since the above minimization is over an integer submodular polyhedron,
the optimal solution is also an integer number. Therefore, minimization problems~\eqref{ch3:primal} and~\eqref{ch3:sub_mod} are equivalent.

Let us denote by $\mcl{R}^{(i)}$ the rate region that corresponds to the optimization problem~\eqref{ch3:primal}.
\begin{align}
&\mcl{R}^{(i)} = \{\mbf{R} \in \mathbb{R}^i~|~R(\mcl{S}\cup \{i\})\leq f_{\beta}(\mcl{S} \cup \{i\}), \nonumber \\
&~~~~~~~~~~~~~~~~~~~~~~\forall \mcl{S} \subseteq \{1,2,\ldots,i-1\} \}.
\label{edm:rate_region}
\end{align}
To solve optimization problem~\eqref{ch3:primal}, we apply the dual subgradient method. First, the Lagrangian function of the problem~\eqref{ch3:primal} is
\begin{align}
\mcl{L}(\mbf{R},\boldsymbol{\lambda}) = R_i + \sum_{k=1}^{i-1} \lambda_k (R_k-R_k^{*}),~~\mbf{R} \in \mcl{R}^{(i)},
\end{align}
where $\lambda_k\geq 0$, $k=1,2,\ldots,i-1$.
Then, the dual function $\delta(\boldsymbol{\lambda})$ equals to
\begin{align}
\delta(\boldsymbol{\lambda}) &= \max_{\mbf{R} \in \mcl{R}^{(i)}} \mcl{L}(\mbf{R},\mbf{\lambda}) \nonumber \\
                      &= \max_{\mbf{R} \in \mcl{R}^{(i)}} \left\{ R_i + \sum_{k=1}^{i-1} \lambda_k R_k\right\} - \sum_{k=1}^{i-1} \lambda_k R_k^{*}. \label{ch3:dual_fcn}
\end{align}
Due to the maximization step in~\eqref{ch3:dual_fcn} over multiple hyper-planes, it immediately follows that $\delta(\boldsymbol{\lambda})$ is a convex function. By the weak duality theorem~\cite{boyd2004convex},
\begin{align}
\delta(\mbf{\lambda})\geq R_i^{*},~~\forall \lambda_k\geq 0,~k=1,2,\ldots,i-1. \label{edm:dual:ineq}
\end{align}
Hence,
\begin{align}
\min_{\boldsymbol{\lambda}} \left\{ \delta(\boldsymbol{\lambda})~|~\lambda_k\geq 0,~k=1,2,\ldots,i-1 \right\} \geq R_i^{*}
\end{align}
Since optimization problem~\eqref{ch3:primal} is linear, there is no duality gap, \emph{i.e.},
\begin{align}
R_i^{*} = \min_{\boldsymbol{\lambda}} \left\{ \delta(\boldsymbol{\lambda})~|~\lambda_k\geq 0,~k=1,2,\ldots,i-1 \right\}. \label{ch3:edm:dual}
\end{align}
To solve optimization problem~\eqref{ch3:edm:dual}, we apply the dual subgradient method~\cite{boyd2003subgradient} as follows.
Starting with a feasible iterate $\lambda_k[0]$, $k=1,2,\ldots,i-1$, w.r.t. the optimization problem~\eqref{ch3:edm:dual}, and the step size $\theta_j$, every subsequent iterate $\lambda_k[j+1]$ for all $k = 1,2,\ldots,i-1$, can be recursively computed  as follows
\begin{align}
\lambda_k[j+1]=\left\{ \lambda_k[j] - \theta_j (\tilde{R}_k[j] - R_k^{*}) \right\}_{+}, \label{edm:dual:update}
\end{align}
where $\tilde{R}_k[j]$ is an optimal solution to the problem
\begin{align}
\max_{\mbf{R} \in \mcl{R}^{(i)}} R_i + \sum_{k=1}^{i-1} \lambda_k[j] R_k. \label{ch3:edm:subopt1}
\end{align}
Note that $\tilde{R}_k[j]$, $k=1,2,\ldots,i-1$, is a derivative of the dual function $\delta(\boldsymbol{\lambda}[j])$.
\begin{lemma}\label{edm:lemma1}
An optimal solution to the problem~\eqref{ch3:edm:subopt1} can be obtained as follows. Let $t(1),t(2),\ldots,t(i-1)$ be an ordering of
$1,2,\ldots,i-1$ such that $\lambda_{t(1)}\geq \lambda_{t(2)}\geq \cdots \geq \lambda_{t(i-1)}$. Then,
\begin{align}
\tilde{R}_i[j] &= \begin{cases}
                 f_{\beta}(\{i\}), & \text{if}~\lambda_{t(1)}\leq 1, \nonumber \\
                 0, & \text{otherwise}. \nonumber
                 \end{cases} \\
\tilde{R}_{t(k)} &= f_{\beta} (\mcl{S}_{t(k)} \cup \{i\}) - \sum_{u=1}^{k-1} \tilde{R}_{t(u)}[j] - \tilde{R}_i[j],
\end{align}
for $k=1,2,\ldots,i-1$, where $\mcl{S}_{t(k)}\triangleq \{t(1),t(2),\ldots,t(k)\}$.
\end{lemma}
Proof of this Lemma is provided in Appendix~\ref{app:edm:lemma1}
\begin{remark}\label{rmk:edm_step:complexity}
The complexity of the algorithm proposed by Lemma~\ref{edm:lemma1} is $\mcl{O}(i\log i + i\cdot N^3)$.
\end{remark}

The reason why we apply subgradient methods instead of a gradient descent is because function $\delta(\boldsymbol{\lambda})[j]$ even though convex, is not differentiable. From Lemma~\ref{edm:lemma1}, it follows that for a given $\boldsymbol{\lambda}[j]$,
there may be more than one maximizer of the problem~\eqref{ch3:edm:subopt1}.
Due to possibility of having more than one direction along which we can update vector $\boldsymbol{\lambda}[j]$ according to~\eqref{edm:dual:update}, subgradient method is not technically a descent method; the function value $\delta(\boldsymbol{\lambda}[j])$ may often increase in the consecutive steps. For that reason, at each step we keep track of the smallest solution up to that point in time
\begin{align}
\tilde{\boldsymbol{\lambda}}[j] = \argmin\{\delta(\boldsymbol{\lambda}[0]),\delta(\boldsymbol{\lambda}[1]),\ldots,\delta(\boldsymbol{\lambda}[j])\}. \label{edm:best}
\end{align}

Before we go any further, note that the primal optimization problem~\eqref{ch3:primal} is over real vectors. However, the minimization~\eqref{ch3:min_sum}
is an integer optimization problem. As pointed out above, the optimal solution of the problem~\eqref{ch3:primal} is equal to the solution of the problem~\eqref{ch3:min_sum}. Therefore, we can choose the number of iterations $l$ of the dual subgradient method such that we get ``close enough''
to an integer solution. In other words,
\begin{align}
\left| \delta(\tilde{\boldsymbol{\lambda}}[l]) - R_i^{*} \right|\leq \varepsilon, \label{edm:condition}
\end{align}
where $\varepsilon<0.5$. Then,
\begin{align}
R_i^{*} = \text{round}(\delta(\tilde{\boldsymbol{\lambda}}[l]) ). \label{edm:optimal}
\end{align}

\subsection*{Convergence Analysis}

In this section we explore the relationship between the number of iterations of the dual subgradient method $l$, and the step size $\theta_j$,
such that it is guaranteed that~\eqref{edm:optimal} provides the optimal solution.
%Following the notes on subgradient methods provided in~\cite{boyd2003subgradient},

\begin{lemma}\label{lm:convergence}
Let $\boldsymbol{\lambda}^{*}$ be an optimal vector that minimizes the dual function $\delta$. Then,
\begin{align}
&~~~\delta(\tilde{\boldsymbol{\lambda}}[l-1]) - \delta(\boldsymbol{\lambda}^{*}) \nonumber \\
&\leq \frac{\left(\sum_{k=1}^{i-1} \lambda_k[0]\right)^2 + \left(\sum_{k=1}^{i-1} \lambda_k^{*}\right)^2 +2N^2\sum_{j=0}^{l-1} \theta_j^2}{2\sum_{j=0}^{l-1} \theta_j}. \label{lm:convergence:ineq}
\end{align}
\end{lemma}
Proof of Lemma~\ref{lm:convergence} can be derived from the notes on subgradient methods presented in~\cite{boyd2003subgradient}. For the sake of completeness, we provide its entire proof in Appendix~\ref{app:lm:convergence}.

Since by Lemma~\ref{lm:convergence}, $\boldsymbol{\lambda}^{*}$ can be an arbitrary minimizer of the dual function $\delta$,
%we only need to show that there exists a bounded dual optimal solution $\boldsymbol{\lambda}^{*}$.
let us choose $\boldsymbol{\lambda}^{*}$ that can be bounded as suggested by the following lemma.
\begin{lemma} \label{lm:primal_bound}
There exists an optimal solution to the problem~\eqref{ch3:edm:dual} that satisfies
\begin{align}
\sum_{k=1}^{i-1} \lambda_k^{*} \leq m. \label{edm:convergence:ineq5}
\end{align}
\end{lemma}
Proof of this Lemma is provided in Appendix~\ref{app:lm:primal_bound}.
Initial feasible $\boldsymbol{\lambda}[0]$ can be chosen as follows
\begin{align}
\lambda_k[0]=0,~~\forall k \in \{1,2,\ldots,i-1\}. \label{edm:convergence:ineq6}
\end{align}
Combining~\eqref{lm:convergence:ineq}, \eqref{edm:convergence:ineq5} and~\eqref{edm:convergence:ineq6}, we obtain
\begin{align}
\delta(\tilde{\boldsymbol{\lambda}}[l-1]) - \delta(\boldsymbol{\lambda}^{*})\leq
\frac{m^2 +2N^2\sum_{j=0}^{l-1} \theta_j^2}{2\sum_{j=0}^{l-1} \theta_j}. \label{edm:convergence:ineq6}
\end{align}

There are many ways to choose the step size that satisfies the condition~\eqref{edm:convergence:ineq6}. Here, we briefly examine the constant step size, where
$\theta_j=\theta$, $j=0,1,2,\ldots$. For other choices on selecting an appropriate step size $\theta_j$, we refer the interested reader to the notes on subgradient
methods. When $\theta_j = \theta$, the inequality~\eqref{edm:convergence:ineq6} becomes
\begin{align}
\delta(\tilde{\boldsymbol{\lambda}}[l-1]) - \delta(\boldsymbol{\lambda}^{*})\leq
\frac{m^2 +2N^2l\theta^2}{2l\theta}. \label{edm:convergence:ineq7}
\end{align}
Hence, the condition~\eqref{edm:condition} is satisfied when
\begin{align}
\frac{m^2 +2N^2l\theta^2}{2l\theta} < \frac{1}{2}. \label{edm:convergence:ineq8}
\end{align}
It can be easily verified that~\eqref{edm:convergence:ineq8} holds when
\begin{align}
\theta &<\frac{1}{2N^2}, \label{edm:convergence:ineq9} \\
l &> \frac{m^2}{\theta(1-2N^2\theta)}. \label{edm:convergence:ineq10}
\end{align}

Putting all these results together, the minimization~\eqref{ch3:sub_mod} can be obtained by running Algorithm~\ref{ch3:alg:edm:noncomb}.
\begin{algorithm}
\caption{Minimization~\eqref{sub_mod} of Algorithm~\ref{alg:modedm}}
\label{ch3:alg:edm:noncomb}
\begin{algorithmic}[1]
\STATE Select parameters $l$, and $\theta_j$, $j=0,1,\ldots,l-1$ such that
       \begin{align}
       \frac{m^2 +2N^2\sum_{j=0}^{l-1} \theta_j^2}{2\sum_{j=0}^{l-1} \theta_j}<\frac{1}{2}.
       \end{align}
\STATE Set $\lambda_k[0]=0$, $k=1,2,\ldots,i-1$, and $\tilde{\boldsymbol{\lambda}}[0]=\boldsymbol{\lambda}[0]$.
\FOR {$j=0$ to $l-1$}
\STATE \begin{align}
        \lambda_k[j+1]=\left\{ \lambda_k[j] - \theta_j (\tilde{R}_k[j] - R_k^{*}) \right\}_{+}, \nonumber
       \end{align}
       for $k=1,2,\ldots,i-1$, where $\tilde{\mbf{R}}[j]$ is computed according to Lemma~\ref{edm:lemma1}.
\STATE \begin{align}
       \tilde{\boldsymbol{\lambda}}[j+1] = \argmin\left\{ \delta(\boldsymbol{\lambda}[j+1]), \delta(\tilde{\boldsymbol{\lambda}}[j]) \right\}. \nonumber
       \end{align}
\ENDFOR
\STATE \begin{align}
       R_i^{*} = \text{round}\left( \delta(\tilde{\boldsymbol{\lambda}}[l]) \right). \nonumber
       \end{align}
\end{algorithmic}
\end{algorithm}
\begin{remark}
From Remark~\ref{rmk:edm_step:complexity} it follows that the complexity of Algorithm~\ref{ch3:alg:edm:noncomb} is
$SFM(m) = \mcl{O}(lm\log m + lm \gamma)$. For a constant step size $\theta$, from~\eqref{edm:convergence:ineq9} and~\eqref{edm:convergence:ineq10} it follows that the complexity of Algorithm~\ref{ch3:alg:edm:noncomb} can be bounded by
$\mcl{O}(N^2m^3 \log m + N^2m^3\gamma)$.
\end{remark}

\subsection{General Convex Separable Cost}\label{sec:generalcost}

In the previous section, for the linear cost function, we applied Edmonds' algorithm in order to obtain the optimal rate allocation.
Edmonds' algorithm is greedy by its nature since all rate updates are reaching the boundary of polyhedron $P(g_{\beta})$.
This effectively means that Edmonds' algorithm provides rate allocations that are vertices of the base polyhedron $B(g_{\beta})$.
While this was an optimal approach in the case of linear objectives, for the general convex separable cost function the optimal
rate vector may not belong to a vertex of $B(g_{\beta})$. We will show this in Example~\ref{example3}.

The general convex cost optimization problem
\begin{align}
\min_{\mbf{R} \in \mathbb{Z}^m} \sum_{i=1}^m \varphi_i(R_i),~~\text{s.t.}~~\mbf{R} \in B(g_{\beta}) \label{convex:objective}
\end{align}
is known as a \emph{resource allocation problem under submodular constraints}~\cite{ibaraki1988resource},
and it can be solved by applying the following intuitive approach: instead of applying greedy scheme, we will incrementally update by one symbol in $\mathbb{F}_q$ a communication rate of a user that has the minimal discrete derivative (see Algorithm~\ref{alg:allocation}).

\begin{algorithm}[h]
\caption{Minimizing convex separable cost under submodular constraints}
\label{alg:allocation}
\begin{algorithmic}[1]
\STATE Set $R_i=0$, $\forall i\in \mcl{M}$
\FOR {$j=1$ to $\beta$}
\STATE Find $i^{*} \in \mcl{M}$ such that
       \begin{align}
       %\hspace{-0.24in}
       i^{*} = \argmin_{i \in \mcl{M}} \left\{ d_i(R_i+1)~|~\mbf{R}+\mbf{e}(i) \in P(g_{\beta})\right\}, \nonumber
       \end{align}
       where $d_i(R_i+1)\triangleq \varphi_i(R_i+1) - \varphi_i(R_i)$, and
       $\mbf{e}(i)$ is the unit basis $m$-dimensional vector with $i^{th}$ coordinate equals to $1$.
\STATE Set $R_{i^{*}} = R_{i^{*}} + 1$.
\ENDFOR
\STATE $\mbf{R}^{*}=\mbf{R}$ is an optimal rate vector w.r.t. the problem~\eqref{convex:objective}.
\end{algorithmic}
\end{algorithm}
\begin{definition}\label{def:transset}
Let us define set $\mcl{T}_j$ to be the set of all users that are in iteration $j$ of Algorithm~\ref{alg:allocation} allowed to update their transmission rates.
\begin{align}
\mcl{T}_j \triangleq \left\{i~|~\mbf{R}+\mbf{e}(i) \in P(g_{\beta})\right\}. \label{transset}
\end{align}
\end{definition}

The question is how to efficiently recover set $\mcl{T}_j$ in each round of Algorithm~\ref{alg:allocation}. First, we observe
that $P(g_{\beta}) = P(f_{\beta})$ according to Theorem~\ref{thm:dilw}. Second, note that in Algorithm~\ref{alg:modedm},
the minimization~\eqref{sub_mod} outputs the maximum rate vector update along one coordinate. Therefore, we only need to verify
whether such update is at least equal to one symbol in $\mathbb{F}_q$. In other words, $i \in \mcl{T}_j$ if
\begin{align}
\min_{\mcl{S} \subseteq \mcl{M} \setminus \{i\}} \{f_{\beta}(\mcl{S} \cup \{i\})-R(\mcl{S}\cup \{i\})\} \geq 1. \label{sub_check}
\end{align}
Putting these results together, we can obtain a polynomial time solution to problem~\eqref{de_problem1} by applying Algorithm~\ref{alg:allocation:intsub}.

\begin{algorithm}[h]
\caption{Minimizing convex separable cost under intersecting submodular constraints}
\label{alg:allocation:intsub}
\begin{algorithmic}[1]
\STATE Set $R_i=0$, $\forall i\in \mcl{M}$
\FOR {$j=1$ to $\beta$}
\STATE Construct set $\mcl{T}_j$ as follows
       \begin{align}
       \hspace{-0.05in}
       \mcl{T}_j = \left\{i:\min_{\mcl{S}\subseteq \mcl{M} \setminus \{i\}} \{f_{\beta}(\mcl{S}\cup \{i\})-R(\mcl{S} \cup \{i\})\} \geq 1 \right\}. \label{Tj:intesub}
       \end{align}
\STATE Find $i^{*} \in \mcl{T}_j$ such that
       \begin{align}
       i^{*} = \argmin_{i \in \mcl{T}_j} \{d_i(R_i+1)\}. \nonumber
       \end{align}
\STATE Set $R_{i^{*}} = R_{i^{*}} + 1$.
\ENDFOR
\STATE $\mbf{R}^{*}=\mbf{R}$ is an optimal rate vector w.r.t. the problem~\eqref{convex:objective}.
\end{algorithmic}
\end{algorithm}

The complexity of~\eqref{sub_check} is $SFM(m)$, since the function $f_{\beta}(\mcl{S})-R^{*}(\mcl{S})$ is fully submodular.
This check can be done either by minimizing submodular function as suggested in~\eqref{sub_check} or by running the dual subgradient
algorithm similar to the one proposed in Section~\ref{sec:subgradients}. Here, we briefly explain the differences.
First, rate region $\mcl{R}^{(i)}$ defined in~\eqref{edm:rate_region}, now has the following form
\begin{align}
&\mcl{R}^{(i)} = \{\mbf{R} \in \mathbb{R}^m~|~R(\mcl{S}\cup \{i\})\leq f_{\beta}(\mcl{S} \cup \{i\}), \label{edm:rate_region1} \\
&~~~~~~~~~~~\forall \mcl{S} \subseteq \mcl{M} \setminus \{i\} \}. \nonumber
\end{align}
Let us denote by $\mbf{R}^{*} \in \mathbb{R}^m$ the current rate allocation in round $j$ of Algorithm~\ref{alg:allocation:intsub}.
Then, if the maximization
\begin{align}
&\max_{\mbf{R} \in \mcl{R}^{(i)}} R_i, \label{ch3:primal:check} \\
&~~~~~\text{s.t.}~R_k\geq R_k^{*},~k=1,2,\ldots,m,
\end{align}
is at least $1$, then $i \in \mcl{T}_j$. Problem~\eqref{ch3:primal:check} can be solved by following the same steps in solving the
dual problem as in Section~\ref{sec:subgradients}.

\begin{remark}
At each iteration, Algorithm~\ref{alg:allocation:intsub} calls~\eqref{Tj:intesub} $m$ times, and there are total of $\beta$ iterations. Therefore,
the complexity of Algorithm~\ref{alg:allocation:intsub} is \mbox{$\mcl{O}(m \cdot \beta \cdot SFM(m))$}.
\end{remark}

\begin{lemma}\label{lm:cost_cap}
Let us denote by $\beta^{*}$ the minimizer of the function $h$ defined in~\eqref{de_problem1}. Then, $\beta^{*}\leq N$.
\end{lemma}
Proof of Lemma~\ref{lm:cost_cap} is provided in Appendix~\ref{app:lm:cost_cap}.

For the general non-decreasing set of functions $\varphi_i$, $i\in \mcl{M}$, from Theorem~\ref{thm:convex} we know that function $h$
is convex. Moreover, by Lemma~\ref{lm:cost_cap} it follows that the minimizer of $h$ is at most equal to $N$. Therefore, in order to minimize $h$, we can apply
Algorithm~\ref{alg:linear_cost} which computes $h(\beta)$ for any $\beta$ by applying Algorithm~\ref{alg:allocation}. Thus, the overall
complexity of the proposed solution is \mbox{$\mcl{O}(m \cdot SFM(m) \cdot N\log N)$}.

\subsection{Fairness under the fixed sum-rate budget} \label{sec:fairness}

In this section we study the problem where for the fixed feasible sum-rate budget $\beta$, the goal is to distribute communication
load to users as evenly as possible. Linear cost function is by its nature ``unfair,'' since it
can potentially result in a communication scheme where only a small group of users transmit packets. For the fixed sum-rate budget, the ``fairness'' can be achieved by introducing an
uniform, non-decreasing (in the integer domain) objective $\varphi_i(R_i) = R_i\log R_i$, $i=1,\ldots,m$, and it is illustrated in the example below.
\begin{example} \label{example3}
Consider the same three-user problem as in Example~\ref{example1}.
\begin{align}
\mbf{x}_1&=\left[
             \begin{array}{cc}
               w_1 & w_2
             \end{array}
           \right]^T, \nonumber \\
\mbf{x}_2&=\left[
             \begin{array}{cccc}
               w_2 & w_4 & w_5 & w_6
             \end{array}
           \right]^T, \nonumber \\
\mbf{x}_3&=\left[
             \begin{array}{cccc}
               w_3 & w_4 & w_5 & w_6
             \end{array}
           \right]^T, \nonumber
\end{align}
where $w_i \in \mathbb{F}_q$, $i=1,\ldots,6$.

\begin{figure}[h]
\begin{center}
\includegraphics[scale=0.55]{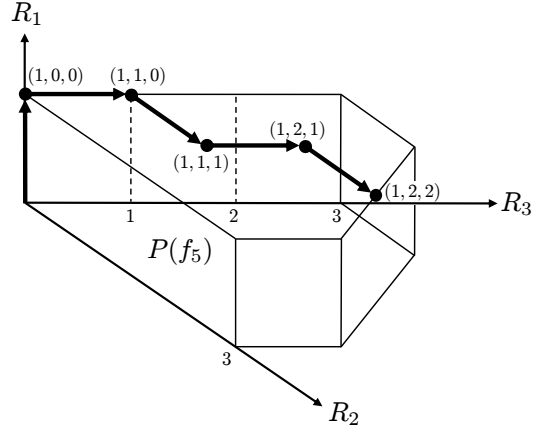}
\end{center}
\vspace{-0.2in}
\caption{Algorithm~\ref{alg:allocation} applied to the three-user problem from Example~\ref{example3}, with the
         cost function $\sum_{i=1}^3 R_i \log R_i$ and the fixed sum-rate $R_1+R_2+R_3=5$. To minimize the cost,
         in each iteration we update the rate of the user who has transmitted the least amount of symbols in $\mathbb{F}_q$
         such that the update still belongs to polyhedron $P(f_{\beta})$.} \label{fig:example3}
\end{figure}

%For a feasible sum-rate $\beta=5$, the intersecting submodular function $f_{\beta}$ defined in~\eqref{dilw1} equals to:
%\begin{align}
%&f_{\beta}(\{1\})=1,~~f_{\beta}(\{2\})=3,~~f_{\beta}(\{3\})=3, \nonumber \\
%&f_{\beta}(\{1,2\})=4,~~f_{\beta}(\{1,3\})=5,~~f_{\beta}(\{2,3\})=4, \nonumber \\
%&f_{\beta}(\{1,2,3\})=5. \label{fcn:fb}
%\end{align}
In case of a linear objective $2R_1+R_2+3R_3$, for a given sum-rate $\beta=5$, we showed hat the optimal \emph{DE}-rate vector,
obtained by using Algorithm~\ref{alg:modedm}, belongs to a vertex of the base polyhedron $B(f_{\beta})$:
\begin{align}
R^{*}_1=1,~~R^{*}_2=3,~~R^{*}_3=1. \label{de_lin_cost}
\end{align}
Let us now analyze the case when the objective is $\varphi_i(R_i)=R_i\log R_i$, $i=1,2,3$.
Following the notation of Algorithm~\ref{alg:allocation}, we have that
\begin{align}
d_i(R_i+1) = (R_i+1)\log(R_i+1) - R_i\log R_i.
\end{align}
It is not hard to show that the above function $d_i(\cdot)$ is increasing. Hence, the minimization step in Algorithm~\ref{alg:allocation:intsub} can be written as
\begin{align}
i^{*} = \argmin_{i \in \mcl{T}_j} R_i, \label{uniform}
\end{align}
where $\mcl{T}_j$ can be computed from~\eqref{sub_check}, and $j=1,\ldots,\beta$ is an iteration of Algorithm~\ref{alg:allocation:intsub}.
The condition~\eqref{uniform} proves that $\varphi_i(R_i)=R_i \log R_i$ is a good measure for fairness, since it is enforcing the transmission
vector $\mbf{R}$ to be as uniform as possible.
The execution steps of Algorithm~\ref{alg:allocation} are shown in Figure~\ref{alg:allocation}. It can be verified that
\begin{align}
\mcl{T}_1=\{1,2,3\},~\mcl{T}_2=\mcl{T}_3=\mcl{T}_4=\mcl{T}_5=\{2,3\}. \nonumber
\end{align}
Therefore, the optimal \emph{DE}-rate vector for this example is $R^{*}_1=1$, $R^{*}_2=2$, $R^{*}_3=2$.
%Observe that unlike the linear cost solution~\eqref{de_lin_cost}, the uniform objective enforces the communication ``budget'' to be
%spread across all the users as evenly as possible.
\end{example}

\section{Code Construction} \label{sec:code}

\begin{figure}[h]
\begin{center}
\includegraphics[scale=0.5]{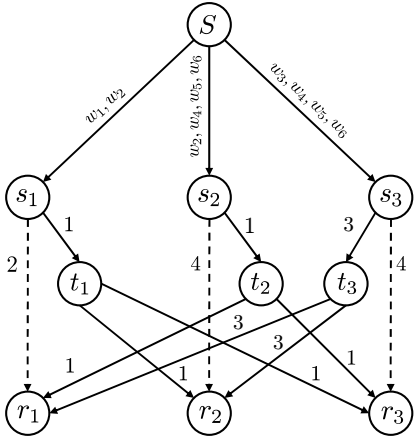}
\end{center}
\vspace{-0.2in}
\caption{Multicast network constructed from the source model and the sum-rate optimal \emph{DE}-rate vector $R^{*}_1=1$, $R^{*}_2=1$,
$R^{*}_3=2$. Hence, in an optimal scheme
users $1$, $2$ and $3$ are transmitting $1$, $1$, and $3$ linear combinations of their own observations in $\mathbb{F}_q$, respectively. Each user receives side-information from ``itself'' (through the links $s_i\rightarrow r_i$, $i=1,2,3$) and from the other users (through the links $t_i\rightarrow r_j$, $i,j\in \{1,2,3\}$, $i\neq j$).} \label{fig:multicast}
\end{figure}

In Theorem~\ref{thm:netcode}, we showed that in order to achieve optimal communication rates, it is sufficient for each
user to transmit the optimal number of linear combinations of its observations.
In this section, we show how to efficiently design the transmission scheme. We explain the code construction
on the three user problem from Example~\ref{example1}, where
\begin{align}
\mbf{x}_1&=\left[
             \begin{array}{cc}
               w_1 & w_2
             \end{array}
           \right], \nonumber \\
\mbf{x}_2&=\left[
             \begin{array}{cccc}
               w_2 & w_4 & w_5 & w_6
             \end{array}
           \right], \nonumber \\
\mbf{x}_3&=\left[
             \begin{array}{cccc}
               w_3 & w_4 & w_5 & w_6
             \end{array}
           \right]. \label{code_construction:source_model}
\end{align}
For the objective function $\min R_1+R_2+R_3$, we showed that
the optimal \emph{DE}-rate vector is $R^{*}_1=1$, $R^{*}_2=1$, and $R^{*}_3=3$. This means that in an optimal scheme
users $1$, $2$ and $3$ transmit $1$, $1$, and $3$ linear combinations of their own observations in $\mathbb{F}_q$, respectively.
We design the coding scheme by first constructing the corresponding multicast network (see Figure~\ref{fig:multicast}).
In this construction, notice that there are several types of nodes. First, there is a super node $S$ that has all the packets.
Each user in the system is a transmitter, while in addition, each user is also a receiver.
To model this, we denote $s_1$, $s_2$ and $s_3$ to be the ``transmitting'' nodes, and $r_1$, $r_2$ and $r_3$ to be the ``receiving'' nodes.
The side-information observed by users $1$, $2$ and $3$ gets directly routed from $s_1$, $s_2$ and $s_3$ to the receivers $r_1$, $r_2$ and $r_3$ through direct edges (dashed edges in Figure~\ref{fig:multicast}). To model the broadcast nature of each transmission, we introduce the
``dummy'' nodes $t_1$, $t_2$ and $t_3$, such that the capacity of the links $(s_i,t_i)$ is the same as link capacity $(t_i,r_j)$, $j\neq i$, and is equal to $R^{*}_i$, $\forall i \in \mcl{M}$.

\begin{figure}[h]
\begin{center}
\includegraphics[scale=0.55]{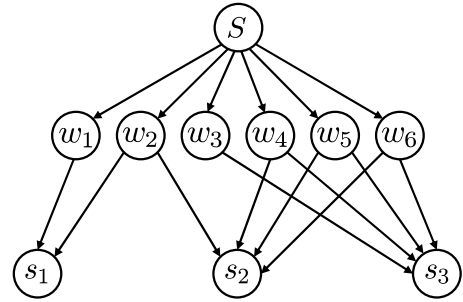}
\end{center}
\vspace{-0.2in}
\caption{When each user observes subset of the file packets, we can model the observations by adding an extra layer of $N=6$ nodes to the graph in Figure~\ref{fig:multicast}. Each extra node represents one file packet, and all extra edges are of capacity $1$. Then, users' observations  can be modeled by connecting nodes from this layer to the users' nodes $s_1$, $s_2$ and $s_3$ according to~\eqref{code_construction:source_model}.} \label{fig:multicast:mod}
\end{figure}

Now, when we have a well-defined network it is only left to figure out transmissions on all the edges.
For instance, this can be achieved using Jaggi et al. algorithm~\cite{Jaggi05}. The first step of this algorithm
is to determine $N=6$ disjoint paths from the super-node $S$ to each receiver $r_1$, $r_2$ and $r_3$ by using the Ford-Fulkerson
algorithm \cite{B98}. Such paths are designed to carry linearly independent messages from the super node to the receivers. When each user observes some subset of the file packets (as it is the case in this example), we can directly apply Jaggi \emph{et al.}
algorithm to this problem by slightly modifying the upper portion of the multicast network from Figure~\ref{fig:multicast} (see Figure~\ref{fig:multicast:mod}).
Note that in this case, we were able to model observations of each user simply by adding one more layer of nodes which represent individual file packets, and then
connecting these packet nodes with each user according to~\eqref{code_construction:source_model}. In other words, the entire source model and the communication model can be represented by multicast acyclic graph. Therefore, Jaggi \emph{et al.} algorithm would find actual transmissions of each user in polynomial time.

In the case of general linear packet model, it is not possible to represent users' observations just by adding one extra layer of nodes to the multicast graph as in Figure~\ref{fig:multicast:mod}.
This is because there is an underlying correlation between all the linear combinations that appear in the users' observation vectors, and it would be suboptimal to treat
all these combinations independently. For that reason, it is more suitable to apply Harvey's algorithm~\cite{H05} which is based
on matrix representation of transmissions in the network~\cite{KM03}, \cite{ho2006random}, and simultaneous matrix completion problem over finite fields. In the remainder of this section, we briefly examine building blocks of this code construction algorithm.

First, we choose source matrix $\mbf{A}$ to be the side-information matrix of all users as,
\begin{align}
\mbf{A}=\left[
          \begin{array}{cccc}
            \mbf{A}_1^T & \ldots & \mbf{A}_m^T & \mbf{0} \ldots \mbf{0} \\
          \end{array}
        \right],
\end{align}
where $\mbf{A}_i$ corresponds to the observation matrix defined in \eqref{model:eq1}.
Matrix $\mbf{A}$ is an $N \times \ell$ matrix, where $\ell$ is the total number of edges in the network.

The transfer matrix $\mbf{M}(r_i)$ from the super-node $S$ to any receiver $r_i$, $i \in \mcl{A}$ can be obtained as shown in~\cite{KM03}.
It is a $N \times N$ matrix with the input vector $\mbf{w}$, and the output vector corresponding to the observations at the receiver $r_i$.
\begin{align}
\mbf{M}(r_i) = \mbf{A} (\mbf{I} - \mbf{\Gamma})^{-1} \mbf{D}(r_i),~~~i=1,2,\ldots,m,
\end{align}
where $\mbf{\Gamma}$ is adjacency matrix of the multicast network,
and $\mbf{D}(r_i)$ is an output matrix. For more details on how these matrices are constructed,
we refer the interested reader to the reference \cite{KM03}.

A multicast problem has a network coding solution if and only if each matrix $\mbf{M}(r_i)$
is non-singular. In \cite{H05}, the author showed that for the \emph{expanded transfer matrix} defined as
\begin{align}
\mbf{E}(r_i)=
\left[
  \begin{array}{cc}
    \mbf{A} & \mbf{0} \\
    \mbf{I}-\mbf{\Gamma} & \mbf{D}(r_i) \\
  \end{array}
\right],~~~i=1,2,\ldots,m,
\end{align}
it holds that $\det (\mbf{M}(r_i))=\pm \det (\mbf{E}(r_i))$.

Some entries of the matrices $\mbf{\Gamma}$ and $\mbf{D}(r_i)$, $i=1,2,\ldots,m$, are
unknowns. To obtain the actual transmissions on all the edges, it is necessary to replace those unknown entries
with elements from $\mathbb{F}_q$ such that all matrices $\mbf{E}(r_i)$, $i=1,2,\ldots,m$, have full rank.
This is known as a simultaneous matrix completion problem and it is solved in \cite{H05} in polynomial time provided that $|\mathbb{F}_q|> m$.
\begin{remark}
The complexity of the algorithm proposed in~\cite{H05} applied to our problem is
\mbox{$\mcl{O}(m^4\cdot \gamma \cdot \log(m\cdot N))$}, where $\gamma$ is the complexity of computing rank.
\end{remark} 
\section{Randomized Algorithm}  \label{sec:rand_alg}

In this section we combine Algorithm~\ref{alg:allocation} with the linear network coding scheme to produce a randomized solution
to the optimization problem~\eqref{de_problem1} of linear complexity (in number of users). First, note that Algorithm~\ref{alg:allocation}
is incremental by its nature, \emph{i.e.}, in each iteration we update the rate vector by one symbol in $\mathbb{F}_q$. Say that user~$i$
updates its rate at round $j$ of Algorithm~\ref{alg:allocation}. Along with the rate update, let user~$i$ transmit an appropriately chosen
linear combination of its observations; using the notation from Section~\ref{sec:model}, we have
\begin{align}
v_i^{(j)} = \mbf{b}_i^{(j)} \cdot \mbf{A}_i\cdot \mbf{w}, \label{trans_vector}
\end{align}
where $\mbf{b}_i^{(j)} \in \mathbb{F}^{\ell_i}_q$, is the vector of coefficients
that lead to the optimal communication scheme. We note that those coefficients are not known a priori; they can be figured out by applying the algorithm
proposed in Section~\ref{sec:code} only after the entire optimal \emph{DE}-rate vector is recovered. For now, let us just assume that we have
access to the vectors $\mbf{b}_i^{(j)}$ for all iterations $j=1,\ldots,\beta$, and for all users $i\in \mcl{M}$ that are scheduled to
update their communication rates. Later, we will use random linear network coding argument to relax these assumptions.

In the expression~\eqref{trans_vector}, let us define $\mbf{u}^{(j)}\in \mathbb{F}_q^N$ as
\begin{align}
\mbf{u}^{(j)} \triangleq \mbf{b}_i^{(j)} \cdot \mbf{A}_i. \label{vector:c}
\end{align}
Then, we can write~\eqref{trans_vector} as
\begin{align}
v_i^{(j)} = \mbf{u}^{(j)} \cdot \mbf{w}. \label{trans_vector1}
\end{align}
By generating transmissions along with the rate updates, we can actually reduce the complexity of verifying whether the rate
vector update still belongs to the polyhedron $P(f_{\beta})$. This result is stated in the following theorem.
\begin{theorem} \label{thm:transset}
Let the set $\mcl{T}_j$ be defined as in~\eqref{transset}. Then,
\begin{align}
&\mcl{T}_j = \{ i \in \mcl{M}~|~\text{rank}\left( \mbf{A}_i \cup \mbf{u}^{(1)} \cup \cdots \cup \mbf{u}^{(j-1)} \right) \nonumber \\
&~~~~~~~~~~~~~~~~~~~~> N-(\beta-j+1)  \}. \label{transmit_set}
\end{align}
\end{theorem}
Proof of Theorem~\ref{thm:transset} is provided in Appendix~\ref{app:thm:transset}.

So far we have assumed that the vectors $\mbf{u}^{(j)}$ are provided to us
deterministically, and that they render optimal communication scheme. However, this assumption is unjustifiable since
we saw in Section~\ref{sec:code} that in order to construct a deterministic communication scheme we need to know optimal
\emph{DE}-rate vector beforehand. To go around this problem we invoke a random linear network coding scheme.
The basic idea behind the random linear network coding argument is that if user~$i$ is scheduled to transmit in round~$j$,
then we can choose vectors $\mbf{b}_i^{(j)}$ in~\eqref{trans_vector} uniformly at random over $\mathbb{F}_q^{\ell_i}$.
The following lemma provides a relationship between probability of generating
optimal transmissions and the field size $q$.
\begin{lemma}\label{lm:fieldsize}
For the random linear network coding scheme, the probability of choosing an optimal sequence of vectors $\mbf{u}^{(j)}$, $j=1,2,\ldots,\beta$,
is at least $(1-\frac{m}{q})^{\beta}$.
%$1-\frac{const}{q}$.
\end{lemma}
The proof of Lemma~\ref{lm:fieldsize} directly follows from~\cite{ho2006random}. The idea is to relate this problem to a multicast problem
as in Section~\ref{sec:code}, while assuming that the optimal rates are given. Then, by randomly generating
transmissions on each link, we obtain the exactly same formulation as in~\cite{ho2006random}.

Putting all these results together, from Algorithm~\ref{alg:allocation} we can devise its Randomized counterpart as follows (see Algorithm~\ref{alg:random}).
\begin{algorithm}
\caption{Randomized Algorithm}
\label{alg:random}
\begin{algorithmic}[1]
\STATE Set $R_i=0$, $\forall i\in \mcl{M}$
\FOR {$j=1$ to $\beta$}
\STATE Determine $\mcl{T}_j$ as defined in~\eqref{transmit_set}.
\STATE Find $i^{*} \in \mcl{T}_j$ such that
       \begin{align}
       i^{*} = \text{argmin} \left\{ d_i(R_i+1)~|~i \in \mcl{T}_j \right\}. \nonumber
       \end{align}
\STATE Let $i^{*}$ transmit, and create a transmission $v_{i^{*}}^{(j)}$ by creating a vector
       $\mbf{b}_{i^{*}}^{(j)}$ uniformly at random over $\mathbb{F}_q^{\ell_{i^{*}}}$.
\STATE Set $R_{i^{*}}=R_{i^{*}}+1$.
\ENDFOR
\STATE $\mbf{R}^{*}=\mbf{R}$ is an optimal rate vector w.r.t. the problem~\eqref{convex:objective}.
\end{algorithmic}
\end{algorithm}

\begin{remark}
The complexity of Algorithm~\ref{alg:random} is \mbox{$\mcl{O}(m\cdot\gamma \cdot N)$}, where $\gamma$ is the
complexity of computing rank.
\end{remark}
\begin{remark}
When $\beta$ is not a feasible sum-rate w.r.t. the optimization problem~\eqref{de_problem1},
then after $\beta$ iterations of Algorithm~\ref{alg:random} there exists a user that
cannot reconstruct all the packets. In other words
\begin{align}
\exists i \in \mcl{M},~\text{s.t.}~~\text{rank}\left( \mbf{A}_i \cup \mbf{u}^{(1)} \cup \cdots \cup \mbf{u}^{(\beta)} \right) <N. \nonumber
\end{align}
\end{remark}

In order to solve the optimization problem~\eqref{de_problem}, we can apply a binary search algorithm similar to Algorithm~\ref{alg:linear_cost}.
Thus, the overall complexity of the proposed algorithm is \mbox{$\mcl{O}(m \cdot \gamma \cdot N\log N)$}.

\begin{example}\label{example4}
Let us consider the same problem as in Example~\ref{example3}:
\begin{align}
\mbf{x}_1&=\left[
             \begin{array}{cc}
               w_1 & w_2
             \end{array}
           \right], \nonumber \\
\mbf{x}_2&=\left[
             \begin{array}{cccc}
               w_2 & w_4 & w_5 & w_6
             \end{array}
           \right], \nonumber \\
\mbf{x}_3&=\left[
             \begin{array}{cccc}
               w_3 & w_4 & w_5 & w_6
             \end{array}
           \right], \nonumber
\end{align}
where $w_i \in \mathbb{F}_q$, $i=1,\ldots,6$, and $q$ is some large prime number.
For the uniform objective $\sum_{i=1}^3 R_i\log R_i$ with a fixed sum-rate $\sum_{i=1}^3 R_i=5$, Algorithm~\ref{alg:random} executes the following steps:
\begin{itemize}
\item Set $R_1=R_2=R_3=0$.
\item \mbox{$j=1:$} Updates of the rate vector $\mbf{R}^{*}$ are selected according to the rule~\eqref{uniform}:
                    \begin{align}
                    \argmin \left\{R_i~|~i \in \mcl{T}_1 = \{1,2,3\} \right\}= \{1,2,3\}, \nonumber
                    \end{align}
                    User~$1$ transmit some random linear combination of its observation, say $v_1^{(1)}=w_1+7w_2$. Set
                    \begin{align}
                    R_1=R_1+1=1. \nonumber
                    \end{align}
\item \mbox{$j=2:$} Vector $\mbf{R}$ is updated according to the rule:
                    \begin{align}
                    \argmin \left\{R_i~|~i\in \mcl{T}_2 = \{2,3\} \right\}=\{3\}. \nonumber
                    \end{align}
                    User~$3$ transmit some random linear combination of its observation, say $v_3^{(2)}=w_3+w_4+5w_5+11w_6$. Set
                    \begin{align}
                    R_3=R_3+1=1. \nonumber
                    \end{align}
\item \mbox{$j=3:$} Vector $\mbf{R}$ is updated according to the rule:
                    \begin{align}
                    \argmin \left\{R_i~|~i\in \mcl{T}_3=\{2,3\} \right\}=\{2\}. \nonumber
                    \end{align}
                    User~$2$ transmit some random linear combination of its observation, say $v_2^{(3)}=4w_2+3w_4+13w_5+8w_6$. Set
                    \begin{align}
                    R_2=R_2+1=1. \nonumber
                    \end{align}
\item \mbox{$j=4:$} Vector $\mbf{R}$ is updated according to the rule:
                    \begin{align}
                    \argmin \left\{R_i~|~i\in \mcl{T}_4=\{2,3\} \right\}=\{3\}. \nonumber
                    \end{align}
                    User~$3$ transmit some random linear combination of its observation, say $v_3^{(4)}=9w_3+5w_4+14w_5+17w_6$. Set
                    \begin{align}
                    R_3=R_3+1=2. \nonumber
                    \end{align}
\item \mbox{$j=5:$} Vector $\mbf{R}$ is updated according to the rule:
                    \begin{align}
                    \argmin \left\{R_i~|~i\in \mcl{T}_5=\{2,3\} \right\}=\{2\}. \nonumber
                    \end{align}
                    User~$2$ transmit some random linear combination of its observation, say $v_2^{(5)}=11w_2+2w_4+18w_5+6w_6$. Set
                    \begin{align}
                    R_2=R_2+1=2. \nonumber
                    \end{align}
\item $\mbf{R}^{*} = \mbf{R}$ is an optimal \emph{DE}-rate vector w.r.t. the uniform objective and the condition $R(\mcl{M})=5$.
\end{itemize}
It can be verified that after this round of communication all the users are able to recover the file.
\end{example}

\section{Introducing Capacity Constraints}\label{sec:cap_constraints}

In this section we explore a data exchange problem where the transmissions of each user can be further restricted.
For instance, we can limit the total number of packets sent by each user.
%This is particularly useful in the scenarios
%where communication consumes a lot of power, and we want to ``spare'' users with low battery.
Say that user ~$i$ is not allowed to transmit more than $c_i$ packets in $\mathbb{F}_q$. Then, optimization problem~\eqref{de_problem} becomes
\begin{align}
\min_{\beta \in \mathbb{Z}_{+}} h(\beta), \label{de_problem_cap}
\end{align}
where $h(\beta)$ can be obtained from~\eqref{de_sum2} by adding capacity constraints.
\begin{align}
&h(\beta) = \min_{\mbf{R} \in \mathbb{Z}^m} \sum_{i=1}^m \varphi_i(R_i),~~\text{s.t.},~~\mbf{R}\in B(g_{\beta}), \label{de_sum4} \\
&~~~~~~~~~~~~~~~~~~~~~~~~~~~~~~~~~~~~~~~R_i\leq c_i,~~\forall i \in \mcl{M}, \nonumber
\end{align}
provided that $g_{\beta}(\mcl{M})=\beta$.
We also assume that the capacity vector $\mbf{c}$ is feasible, \emph{i.e.}, there exists a rate vector $\mbf{R}\in B(g_{\beta})$
such that the capacity constraints in ~\eqref{de_sum4} are satisfied.

In Section~\ref{sec:det_algorithm} we pointed out that the optimality of all the algorithms we studied is guaranteed due to
the fact that the constraint set of the problem~\eqref{de_problem1} constitutes a base polyhedron of a submodular function.
In this section we show that by adding individual capacity constraints, the constraint
set in~\eqref{de_problem1} also forms a base polyhedron of a submodular function. This implies that in such a case we can
still apply every algorithm developed so far in order to obtain an optimal \emph{DE}-rate vector.

We begin our analysis by defining the restriction of a submodular function (see~\cite{ibaraki1988resource} for the reference).
\begin{definition}\label{def:restriction}
For a submodular function $g_{\beta}:2^{\mcl{M}}\rightarrow \mathbb{Z}$, and a vector $\mbf{c} \in \mathbb{Z}^m$, define a function
$g_{\beta}^{\mbf{c}}:2^{\mcl{M}}\rightarrow \mathbb{Z}$ by
\begin{align}
g_{\beta}^{\mbf{c}}(\mcl{S}) \triangleq \min_{\mcl{V}} \left\{ g_{\beta}(\mcl{V})+c(\mcl{S} \setminus \mcl{V})~|~\mcl{V}\subseteq \mcl{S} \right\},~~\forall \mcl{S}\subseteq\mcl{M}. \label{restriction}
\end{align}
The submodular function $g_{\beta}^{\mbf{c}}$ is called the \emph{restriction of $g_{\beta}$ by vector $\mbf{c}$}.
\end{definition}
\begin{theorem}[Theorem 8.2.1 in~\cite{ibaraki1988resource}]
Let $g_{\beta}^{\mbf{c}}$ be restriction of a submodular function $g_{\beta}$ by vector $\mbf{c}$. Then, $g_{\beta}^{\mbf{c}}$ is submodular.
\end{theorem}

\begin{theorem} \label{therem:restriction}
For a submodular function $g_{\beta}$ defined in~\eqref{dilw1} and a feasible capacity vector $\mbf{c}$ w.r.t. problem~\eqref{de_sum4}, the base polyhedron $B(g_{\beta}^{\mbf{c}})$ of the restriction of $g_{\beta}$ by $\mbf{c}$, is given by
\begin{align}
B(g_{\beta}^{\mbf{c}}) = \left\{ \mbf{R}~|~\mbf{R}\in B(g_{\beta}),~~R_i\leq c_i,~\forall i\in \mcl{M} \right\}, \label{thm:identity}
\end{align}
provided that the sum-rate $\beta$ and the capacity vector $\mbf{c}$ are feasible w.r.t. the optimization problem~\eqref{de_problem1}.
\end{theorem}
Proof of Theorem~\ref{therem:restriction} is provided in Appendix~\ref{app:theorem:restrictions}.

From Theorem~\ref{therem:restriction} it follows that the constraint set of~\eqref{de_sum4} forms a submodular polyhedron $B(g_{\beta}^{\mbf{c}})$,
which further implies that all the algorithms developed so far can be applied to obtain an optimal \emph{DE}-rate vector. For instance,
with capacity constraints, Step~4 of Algorithm~\ref{alg:edm} becomes
\begin{align}
&R^{*}_{j(i)} = \min\{c_{j(i)},g_{\beta}(\{j(1),j(2),\ldots,j(i)\}) \label{edm:cap} \\
&~~~~~~~~~~~~~~~~~~~- g_{\beta}(\{j(1),j(2),\ldots,j(i-1)\})\}. \nonumber
\end{align}
This modification propagates to Algorithm~\ref{alg:modedm} as well.
Similarly, at iteration~$j$, Step~4 of Algorithm~\ref{alg:allocation:intsub} and Step~3 of Algorithm~\ref{alg:random}  is modified as follows
\begin{align}
&i^{*} = \text{argmin} \{ d_i(R^{*}_i+1)~|~i \in \mcl{T}_j,~\text{s.t.},~R^{*}_i+1\leq c_i \}.
\end{align}
\begin{remark}
If the capacity vector $\mbf{c}$ was not feasible w.r.t. problem~\eqref{de_sum4}, then all algorithms considered so
far would terminate before reaching sum-rate equal to $\beta$.
\end{remark}

\begin{example}
Let us consider the same problem as in Example~\ref{example1}
\begin{align}
\mbf{x}_1&=\left[
             \begin{array}{cc}
               w_1 & w_2
             \end{array}
           \right], \nonumber \\
\mbf{x}_2&=\left[
             \begin{array}{cccc}
               w_2 & w_4 & w_5 & w_6
             \end{array}
           \right], \nonumber \\
\mbf{x}_3&=\left[
             \begin{array}{cccc}
               w_3 & w_4 & w_5 & w_6
             \end{array}
           \right], \nonumber
\end{align}
where $w_i \in \mathbb{F}_q$. Let the cost function be $R_1+3R_2+2R_3$, the sum-rate $\beta=5$, and the capacity constraints $c_i\leq 2$, $i=1,2,3$.
Then, by applying Algorithm~\ref{alg:modedm} with the modification~\eqref{edm:cap}, we obtain the following result.
\begin{align}
&R^{*}_1 = \min\{f_5(\{1\}),c_1\}=1, \nonumber \\
&R^{*}_3 = \min\left\{\min\{f_5(\{1,3\})-R^{*}_1,f_5(\{2\})\},c_3\right\} = 2, \nonumber \\
&R^{*}_2 = \min\{\min\{f_5(\{1,2,3\})-R^{*}_1-R^{*}_3,f_5(\{1,3\})-R^{*}_1, \nonumber \\
&~~~~~~~~~~~~~~~~~~~~~~~f_5(\{2,3\})-R^{*}_3,f_5(\{2\})\},c_2\} = 2. \nonumber
\end{align}
Without capacity constraints, as it was the case in Example~\ref{example1}, user~$3$ would transmit $3$ packets in $\mathbb{F}_q$.
\end{example} 
\section{Conclusion} \label{sec:conclusion}
In this work we addressed the problem of the data exchange, where each user has some side-information about the file,
and is interested in recovering it. We assumed that the users are allowed to ``talk'' to each other over a noiseless broadcast channel.
For the case when the side information is in the form of the linearly coded packets,
we provided deterministic and randomized polynomial time algorithms for finding an optimal communication scheme, w.r.t.
a separable convex communication cost, that delivers the file to all users.
%We also provide efficient methods for constructing a communication scheme based on the algebraic network coding approach.

%\input{linear_example.tex}
%\input{det_algorithm.tex} \label{sec:alg}
%\input{Intro.tex}
%\input{Model.tex}
%\input{linear_example.tex}
%\input{det_algorithm.tex} \label{sec:alg}
%\input{Conclusion.tex} \label{sec:conclusion}
%\input{appendix.tex} \label{sec:app}
%\input{Connection_to_KeyAgreement.tex}
%\input{Deterministic.tex}
%\input{Appendix.tex}
\appendices
\section{Proof of Theorem~\ref{thm:netcode}}\label{app:thm:netcode}
In order for each user in $\mcl{M}$ to reconstruct the file, it is necessary for all of them to receive a sufficient number
of linear combinations over $\mathbb{F}_q$ so that the observation rank of each user is full.
For instance,  in order for user~$1$ to recover all $N$ packets of the file, it is sufficient for him to select $N-\ell_1$ linear
equations from the remaining $m-1$ users. In this case, user~$2$ can send to user~$1$
\begin{align}
R_2=\text{rank}\left(\mbf{A}_{\{1,2\}} \right)-\text{rank}\left(\mbf{A}_{\{1\}} \right) \label{netcode:eq1}
\end{align}
of its linear equations, after which user~$1$ will have observation rank $\text{rank}\left(\mbf{A}_{\{1,2\}} \right)$. Following this procedure, we
have that the number of linear equations sent by the remaining users is
\begin{align}
R_3 &=\text{rank}\left(\mbf{A}_{\{1,2,3\}} \right)-\text{rank}\left(\mbf{A}_{\{1,2\}} \right) \label{netcode:eq2} \\
    &~~\vdots \nonumber \\
R_m &=\text{rank}\left(\mbf{A}_{\mcl{M}} \right)-\text{rank}\left(\mbf{A}_{\mcl{M} \setminus \{m\}} \right) \nonumber \\
    &= N-\text{rank}\left(\mbf{A}_{\mcl{M} \setminus \{m\}} \right). \label{netcode:eq3}
\end{align}
Observe that the number of linear equations each user sends depends upon the ordering of users in equations~\eqref{netcode:eq1} through~\eqref{netcode:eq3}.
Let $j(2),\ldots,j(m)$ be any ordering of $2,\ldots,m$. Then, by applying the same approach as above, we obtain other feasible rate tuples.
\begin{align}
R_{j(2)}&=\text{rank}\left(\mbf{A}_{\{1,j(2)\}} \right)-\text{rank}\left(\mbf{A}_{\{1\}} \right) \label{netcode:eq4} \\
R_{j(3)} &=\text{rank}\left(\mbf{A}_{\{1,j(2),j(3)\}} \right)-\text{rank}\left(\mbf{A}_{\{1,j(2)\}} \right) \label{netcode:eq5} \\
&~~\vdots \nonumber \\
R_{j(m)} &= N-\text{rank}\left(\mbf{A}_{\mcl{M} \setminus \{j(m)\}} \right). \label{netcode:eq6}
\end{align}
From~\eqref{netcode:eq4}-\eqref{netcode:eq6}, observe that
\begin{align}
\sum_{i=t}^{m} R_{j(i)} = N - \text{rank}\left(\mbf{A}_{\{1,j(2),\ldots,j(t-1)\}} \right),~~t=2,\ldots,m. \nonumber
\end{align}
By using this method of ordering, we can reconstruct any vertex of the region
\begin{align}
\sum_{i=t}^{m} R_{j(i)} \geq N - \text{rank}\left(\mbf{A}_{\{1,j(2),\ldots,j(t-1)\}} \right),~~t=2,\ldots,m, \nonumber \\
\text{for all permutations $j(2),\ldots,j(m)$ of the set $\mcl{M} \setminus \{1\}$.}\label{netcode:eq7}
\end{align}
The region in~\eqref{netcode:eq7} is equivalent to
\begin{align}
&\sum_{i \in \mcl{S}} R_i \geq N - \text{rank}\left(\mbf{A}_{\mcl{M} \setminus \mcl{S}}\right),~~
\forall \mcl{S} \subseteq \mcl{M}~~\text{s.t.}~~\{1\} \notin \mcl{S}. \nonumber
\end{align}
Let us denote the above region by $\mcl{R}_1$. Similarly, for users $2$ through $m$, we can define regions $\mcl{R}_2,\ldots,\mcl{R}_m$.
Let us denote by $\mcl{R}_{int}$ the set of all integer vectors $\mathbb{Z}^m$ that belong to the cut-set 
region $\mcl{R}$ defined in~\eqref{cut_set}. Then, it is not hard to show that
%It is not hard to show that the cut-set region $\mcl{R}$ defined in~\eqref{cut_set}, can be expressed as
\begin{align}
\mcl{R}_{int} = \mcl{R}_1 \cap \mcl{R}_2 \cap \cdots \cap \mcl{R}_m.
\end{align}
From the discussion above, we know that if $\mbf{R} \in \mcl{R}_{int}$,
then it is sufficient for user~$i$ to send $R_i$ linear equations separately to all the users, which makes the total
of $(m-1)R_i$ equations over $\mathbb{F}_q$ sent by user~$i$. The key property of the linear network codes is that there exists one set of $R_i$ linear equations that user~$i$ can broadcast and simultaneously satisfy demands of all the remaining users in $\mcl{M}$, provided that the field size $|\mathbb{F}_{q}|$ is large enough~\cite{ACLY00}. Hence, every rate tuple that belongs to $\mcl{R}_{int}$ can be achieved via linear network coding.

\section{Proof of Theorem~\ref{thm:convex}}\label{app:thm:convex}
Consider two feasible sum-rates $\beta_1$ and $\beta_2$ w.r.t. the problem~\eqref{de_problem1}. We show that for any $\lambda\in [0,1]$ 
such that $\lambda\beta_1+(1-\lambda)\beta_2 \in \mathbb{Z}_{+}$ it holds that
$h(\lambda\beta_1+(1-\lambda)\beta_2)\leq \lambda h(\beta_1)+(1-\lambda) h(\beta_2)$. Let
$\mbf{R}^{(1)}$ and $\mbf{R}^{(2)}$ be the optimal rate tuples w.r.t. $h(\beta_1)$ and
$h(\beta_2)$, respectively. Note that
\begin{align}
&\lambda h(\beta_1)+(1-\lambda) h(\beta_2) \nonumber \\
&= \sum_{i=1}^m \left( \lambda\varphi_i(R_i^{(1)})+(1-\lambda)\varphi_i(R_i^{(2)}) \right) \nonumber \\
&\overset{(a)}{\geq} \sum_{i=1}^m \varphi_i(\lambda R_i^{(1)}+(1-\lambda)R_i^{(2)}) =
\sum_{i=1}^m \varphi_i(R^{(\lambda)}_i), \label{lmc:0}
\end{align}
where~(a) follows from the convexity of $\varphi_i$, $\forall i \in \mcl{M}$, and
$\mbf{R}^{(\lambda)} \triangleq \lambda \mbf{R}^{(1)}+(1-\lambda)\mbf{R}^{(2)}$.
Now, we show that $\mbf{R}^{(\lambda)}$ is a feasible \emph{DE}-rate vector for the problem~\eqref{de_problem1} when $\beta = \lambda\beta_1+(1-\lambda) \beta_2$.

Since $R^{(1)}(\mcl{M})=\beta_1$ and $R^{(2)}(\mcl{M})=\beta_2$, it follows that
\begin{align}
R^{(\lambda)}(\mcl{M})&=\lambda R^{(1)}(\mcl{M})+(1-\lambda) R^{(2)}(\mcl{M}) \nonumber \\
&=\lambda\beta_1+(1-\lambda) \beta_2. \label{lmc:1}
\end{align}
Since
\begin{align}
R^{(i)}(\mcl{S})\geq N-\text{rank}(\mbf{A}_{\mcl{M} \setminus \mcl{S}}),~\forall \mcl{S} \subset \mcl{M},~i=1,2, \nonumber
\end{align}
we have
\begin{align}
R^{(\lambda)}(\mcl{S})&=\lambda R^{(1)}(\mcl{S}) + (1-\lambda) R^{(2)}(\mcl{S}) \nonumber \\
&\geq N-\text{rank}(\mbf{A}_{\mcl{M} \setminus \mcl{S}}),~~\forall \mcl{S} \subset \mcl{M}. \label{lmc:2}
\end{align}
From \eqref{lmc:1} and \eqref{lmc:2} it follows that $\mbf{R}^{(\lambda)}$ is a feasible \emph{DE}-rate vector
w.r.t. optimization problem~\eqref{de_problem1} when $\beta=\lambda\beta_1+(1-\lambda)\beta_2$. Therefore,
\mbox{$\sum_{i=1}^m \varphi_i(R^{(\lambda)}_i) \geq h(\lambda\beta_1+(1-\lambda)\beta_2)$}. Hence, from~\eqref{lmc:0}, it follows that
\begin{align}
h(\lambda\beta_1+(1-\lambda)\beta_2)\leq \lambda h(\beta_1)+(1-\lambda) h(\beta_2),
\end{align}
which completes the proof.

\section{Proof of Lemma \ref{lm:f}}
\label{app:lm:f}
When $\mcl{S}\cap \mcl{T}\neq \emptyset$, the following inequality holds due to the submodularity of the rank function
\begin{align}
&f_{\beta}(\mcl{S})+f_{\beta}(\mcl{T}) \nonumber \\
&~~= \text{rank}(\mbf{A}_{\mcl{S}})+\text{rank}(\mbf{A}_{\mcl{T}})-2(N-\beta) \nonumber  \\
&~~\geq \text{rank}(\mbf{A}_{\mcl{S}\cup \mcl{T}})+\text{rank}(\mbf{A}_{\mcl{S}\cap \mcl{T}})-2(N-\beta) \nonumber \\
&~~= f_{\beta}(\mcl{S}\cup \mcl{T}) + f_{\beta}(\mcl{S}\cap \mcl{T}). \label{submod}
\end{align}
To show that the function $f_{\beta}$ is submodular
when $\beta\geq N$, it is only left to consider the case $\mcl{S}\cap \mcl{T} = \emptyset$. Since $f_{\beta}(\emptyset)=0$, we have
\begin{align}
&f_{\beta}(\mcl{S})+f_{\beta}(\mcl{T}) \nonumber \\
&~~=\text{rank}(\mbf{A}_{\mcl{S}})+\text{rank}(\mbf{A}_{\mcl{T}})-2(N-\beta) \nonumber  \\
&~~\geq \text{rank}(\mbf{A}_{\mcl{S}\cup \mcl{T}})-(N-\beta)= f_{\beta}(\mcl{S}\cup \mcl{T}). \label{submod1}
\end{align}
The inequality in~\eqref{submod1} directly follows from the submodularity of the rank function.
\begin{align}
&\text{rank}(\mbf{A}_{\mcl{S}})+\text{rank}(\mbf{A}_{\mcl{T}})-\text{rank}(\mbf{A}_{\mcl{S}\cup \mcl{T}}) \geq 0 \geq \beta - N. \nonumber
\end{align}
This completes the proof.

\section{Proof of Lemma~\ref{edm:lemma1}}\label{app:edm:lemma1}

Let us construct the set function $y:2^{\{1,2,\ldots,i\}} \rightarrow \mathbb{Z}$ as follows
\begin{align}
y(\mcl{S})=\begin{cases}
           0 & \text{if}~\mcl{S}=\emptyset, \nonumber \\
           f_{\beta}(\mcl{S}) & \text{if}~i\in \mcl{S}, \nonumber \\
           f_{\beta}(\mcl{S} \cup \{i\}) & \text{if}~i \notin \mcl{S}. \nonumber
           \end{cases}
\end{align}
First, we show that $\mcl{R}^{(i)}=P(y)$. Let $\mbf{R} \in P(y)$. Then, for any $\mcl{S}\subseteq\{1,2,\ldots,i-1\}$, it follows that
\begin{align}
R(\mcl{S} \cup \{i\})\leq y(\mcl{S}\cup \{i\}) = f_{\beta}(\mcl{S}\cup \{i\}).
\end{align}
Therefore, $\mbf{R} \in \mcl{R}^{(i)}$.

Now, let  $\mbf{R} \in \mcl{R}^{(i)}$. From~\eqref{edm:rate_region} we have
\begin{align}
&R(\mcl{S}\cup \{i\})\leq f_{\beta}(\mcl{S}\cup \{i\}) = y(\mcl{S}\cup \{i\}), \label{edm:lm1:ineq1} \\
&~~~~~~~~~\forall \mcl{S}\subseteq \{1,2,\ldots,i-1\}. \nonumber
\end{align}
Since the rate vector is positive, \eqref{edm:lm1:ineq1} implies that
\begin{align}
R(\mcl{S})\leq f_{\beta}(\mcl{S} \cup \{i\}) = y(\mcl{S}),~~\forall \mcl{S}\subseteq \{1,2,\ldots,i-1\}. \label{edm:lm1:ineq2}
\end{align}
From~\eqref{edm:lm1:ineq1} and~\eqref{edm:lm1:ineq2} it follows that $\mbf{R}\in P(Y)$. Hence, $\mcl{R}^{(i)} = P(Y)$.

%Let $\mcl{S}\subseteq\{1,2,\ldots,i\}$ be such that $i\in \mcl{S}$. Then, for any $\mcl{R}\in P(Y)$
%it holds that
%\begin{align}
%R(\mcl{S})&\leq y(\mcl{S})=f_{\beta}(\mcl{S}\cup \{i\}), \label{edm:lm1:ineq1} \\
%R(\mcl{S}\cup \{i\}) &\leq y(\mcl{S}\cup\{i\})=f_{\beta}(\mcl{S}\cup \{i\}). \label{edm:lm1:ineq2}
%\end{align}
%Since the inequality~\eqref{edm:lm1:ineq1} is redundant compared to the inequality~\eqref{edm:lm1:ineq2}, it immediately follows that
%$\mcl{R}^{(i)}=P(Y)$.

Next, we show that function $y$ is fully submodular. For any $\mcl{S},\mcl{T}\subseteq \{1,2,\ldots,i\}$, let us consider the following 3 cases

\textbf{Case 1:} $i\in \mcl{S}$, $i \notin \mcl{T}$
                        \begin{align}
                        y(\mcl{S})+y(\mcl{T})&=f_{\beta}(\mcl{S})+f_{\beta}(\mcl{T}\cup\{i\}) \nonumber \\
                                             &\overset{(a)}{\geq} f_{\beta}(\mcl{S}\cup \mcl{T}) + f_{\beta}((\mcl{S}\cap \mcl{T}) \cup \{i\}) \nonumber \\
                                             &=y(\mcl{S}\cup\mcl{T})+y(\mcl{S}\cap\mcl{T}), \nonumber
                        \end{align}
                        where (a) is due to intersecting submodularity of function $f_{\beta}$.

\textbf{Case 2:} $i\notin \mcl{S}$, $i \notin \mcl{T}$
                        \begin{align}
                        y(\mcl{S})+y(\mcl{T})&=f_{\beta}(\mcl{S} \cup\{i\})+f_{\beta}(\mcl{T}\cup\{i\}) \nonumber \\
                                             &\geq f_{\beta}(\mcl{S}\cup \mcl{T}\cup\{i\}) + f_{\beta}((\mcl{S}\cap \mcl{T}) \cup \{i\}) \nonumber \\
                                             &=y(\mcl{S}\cup\mcl{T})+y(\mcl{S}\cap\mcl{T}). \nonumber
                        \end{align}

\textbf{Case 3:} $i\in \mcl{S}$, $i \in \mcl{T}$
                        \begin{align}
                        y(\mcl{S})+y(\mcl{T})&=f_{\beta}(\mcl{S})+f_{\beta}(\mcl{T}) \nonumber \\
                                             &\geq f_{\beta}(\mcl{S}\cup \mcl{T}) + f_{\beta}(\mcl{S}\cap \mcl{T}) \nonumber \\
                                             &=y(\mcl{S}\cup\mcl{T})+y(\mcl{S}\cap\mcl{T}). \nonumber
                        \end{align}
Therefore, function $y$ is indeed fully submodular. Hence, problem~\eqref{ch3:edm:subopt1} is a linear optimization problem over a submodular polyhedron, and it can be solved by applying algorithm similar to Algorithm~\ref{alg:edm} (see reference~\cite{E70}).
The only difference is that in this case, the weights in the Step~1 of Algorithm~\ref{alg:edm} should be ordered in a non-increasing order.

If $\lambda_{t(1)}\leq 1$, then
\begin{align}
\tilde{R}_i[j] &= y(\{i\}) = f_{\beta}(\{i\}),\nonumber \\
\tilde{R}_{t(k)}[j]& = y(\mcl{S}_{t(k)}\cup\{i\})-y(\mcl{S}_{t(k-1)} \cup\{i\}) \nonumber \\
                &=f_{\beta} (\mcl{S}_{t(k)} \cup \{i\}) - \sum_{u=1}^{k-1} \tilde{R}_{t(u)}[j] - \tilde{R}_i[j],
\end{align}
for $k=1,2,\ldots,i-1$.

If for some $r\in \{1,2,\ldots,i-1\}$, $\lambda_{t(r)}\geq 1 \geq \lambda_{t(r+1)}$, then
\begin{align}
\tilde{R}_i[j] &= y(\mcl{S}_{t(r)}\cup \{i\}) - y(\mcl{S}_{t(r)}) = 0, \nonumber \\
\tilde{R}_{t(k)}[j]& = y(\mcl{S}_{t(k)}\cup\{i\})-y(\mcl{S}_{t(k-1)} \cup\{i\}) \nonumber \\
                &=f_{\beta} (\mcl{S}_{t(k)} \cup \{i\}) - \sum_{u=1}^{k-1} \tilde{R}_{t(u)}[j],
\end{align}
for $k=1,2,\ldots,i-1$.
This completes the proof of this lemma.

\section{Proof of Lemma~\ref{lm:convergence}}\label{app:lm:convergence}

After $j+1$ iterations of the subgradient algorithm, the Euclidian distance between $\boldsymbol{\lambda}[j+1]$ and 
a minimizer $\boldsymbol{\lambda}^{*}$ of the dual function $\delta$, can be bounded as follows
The distance from 
\begin{align}
&~~~~\sum_{k=1}^{i-1} (\lambda_k[j+1] - \lambda_k^{*})^2 \nonumber \\
&= \sum_{k=1}^{i-1}  \left(\left\{ \lambda_k[j]-\theta_{j}(\tilde{R}_k[j]-R_k^{*}) \right\}_{+}-\lambda_k^{*}\right)^2 \nonumber \\
&\leq \sum_{k=1}^{i-1}  \left(\lambda_k[j]-\theta_{j}(\tilde{R}_k[j]-R_k^{*}) - \lambda_k^{*}\right)^2 \nonumber \\
&= \sum_{k=1}^{i-1} \left( \lambda_k[j] - \lambda_k^{*} \right)^2 - 2\theta_{j}\sum_{k=1}^{i-1} (\tilde{R}_k[j]-R_k^{*})(\lambda_k[j]-\lambda_k^{*}) \nonumber \\ &~~+\theta_{j}^2\sum_{k=1}^{i-1} \left( \tilde{R}_k[j]-R_k^{*} \right)^2 \nonumber \\
&\leq \sum_{k=1}^{i-1} \left( \lambda_k[j] - \lambda_k^{*} \right)^2 - 2\theta_{j}\left(\delta(\boldsymbol{\lambda}[j])-\delta(\boldsymbol{\lambda}^{*})\right) \nonumber \\
&~~+\theta_{j}^2\sum_{k=1}^{i-1} \left( \tilde{R}_k[j]-R_k^{*} \right)^2, \label{edm:convergence:ineq1}
\end{align}
where the last inequality is due to convexity of function $\delta(\boldsymbol{\lambda})$, \emph{i.e.},
\begin{align}
\delta(\boldsymbol{\lambda}[j])-\delta(\boldsymbol{\lambda}^{*}) \leq \sum_{k=1}^{i-1} (\tilde{R}_k[j]-R_k^{*})(\lambda_k[j]-\lambda_k^{*}),
\end{align}
since $\tilde{R}_k[j]-R_k^{*}$ is a partial derivative of $\delta(\boldsymbol{\lambda}[j])$ at coordinate $\lambda_k[j]$, $k=1,2,\ldots,i-1$.
Summing both sides of inequality~\eqref{edm:convergence:ineq1} over $j$ from $0$ to $l-1$, we obtain
\begin{align}
&~~~\sum_{k=1}^{i-1} (\lambda_k[l] - \lambda_k^{*})^2 \nonumber \\
&\leq \sum_{k=1}^{i-1} (\lambda_k[0] - \lambda_k^{*})^2 - 2\sum_{j=0}^{l-1} \theta_j \left(\delta(\boldsymbol{\lambda}[j])-\delta(\boldsymbol{\lambda}^{*})\right) \nonumber \\
&~~+\sum_{j=0}^{l-1} \theta_j^2\sum_{k=1}^{i-1} \left( \tilde{R}_k[j]-R_k^{*} \right)^2.
\end{align}
Therefore,
\begin{align}
&~~2\sum_{j=0}^{l-1} \theta_j \left(\delta(\boldsymbol{\lambda}[j])-\delta(\boldsymbol{\lambda}^{*})\right) \nonumber \\
&\leq \sum_{k=1}^{i-1} (\lambda_k[0] - \lambda_k^{*})^2
+\sum_{j=0}^{l-1} \theta_j^2\sum_{k=1}^{i-1} \left( \tilde{R}_k[j]-R_k^{*} \right)^2. \label{edm:convergence:ineq2}
\end{align}
Since,
\begin{align}
&~~~\sum_{j=0}^{l-1} \theta_j \left(\delta(\boldsymbol{\lambda}[j])-\delta(\boldsymbol{\lambda}^{*})\right)  \nonumber \\
&\geq \sum_{j=0}^l \theta_j
\min_{j\in \{0,1,\ldots,l-1\}} \left(\delta(\boldsymbol{\lambda}[j])-\delta(\boldsymbol{\lambda}^{*})\right),
\end{align}

from~\eqref{edm:convergence:ineq2} and~\eqref{edm:best} we obtain
\begin{align}
&~~\delta(\tilde{\boldsymbol{\lambda}}[l-1]) - \delta(\boldsymbol{\lambda}^{*}) = \min_{j\in \{0,1,\ldots,l-1\}} \delta(\boldsymbol{\lambda}[j])-\delta(\boldsymbol{\lambda}^{*}) \nonumber \\
&\leq \frac{\sum_{k=1}^{i-1} (\lambda_k[0] - \lambda_k^{*})^2 +\sum_{j=0}^{l-1} \theta_j^2\sum_{k=1}^{i-1} \left( \tilde{R}_k[j]-R_k^{*} \right)^2}
{2\sum_{j=0}^{l-1} \theta_j}\nonumber \\
&\leq \frac{\sum_{k=1}^{i-1} (\lambda_k[0] - \lambda_k^{*})^2}{2\sum_{j=0}^{l-1} \theta_j} \nonumber \\
&~~+\frac{\sum_{j=0}^{l-1} \theta_j^2 \left( \sum_{k=1}^{i-1}\left(\tilde{R}_k[j]\right)^2 + \sum_{k=1}^{i-1}\left(R_k^{*}\right)^2 \right)}{2\sum_{j=0}^{l-1} \theta_j} \nonumber \\
&\leq \frac{\sum_{k=1}^{i-1} (\lambda_k[0] - \lambda_k^{*})^2}{2\sum_{j=0}^{l-1} \theta_j} \nonumber \\
&~~+\frac{\sum_{j=0}^{l-1} \theta_j^2 \left( \left(\sum_{k=1}^{i-1}\tilde{R}_k[j]\right)^2 + \left(\sum_{k=1}^{i-1}R_k^{*}\right)^2 \right)}{2\sum_{j=0}^{l-1} \theta_j} \nonumber \\
&\leq \frac{\sum_{k=1}^{i-1} (\lambda_k[0] - \lambda_k^{*})^2 +2N^2\sum_{j=0}^{l-1} \theta_j^2}{2\sum_{j=0}^{l-1} \theta_j}, \label{edm:convergence:ineq3}
\end{align}
where the last inequality holds because $R(\mcl{M})\leq f_{\beta}(\mcl{M})\leq N$ for any achievable \emph{DE}-rate vector $\mbf{R}$.
Continuing with~\eqref{edm:convergence:ineq3}, we have
\begin{align}
&~~~\delta(\tilde{\boldsymbol{\lambda}}[l-1]) - \delta(\boldsymbol{\lambda}^{*}) \nonumber \\
&\leq \frac{\left(\sum_{k=1}^{i-1} \lambda_k[0]\right)^2 + \left(\sum_{k=1}^{i-1} \lambda_k^{*}\right)^2 +2N^2\sum_{j=0}^{l-1} \theta_j^2}{2\sum_{j=0}^{l-1} \theta_j}. \label{edm:convergence:ineq4}
\end{align}
\section{Proof of Lemma~\ref{lm:primal_bound}} \label{app:lm:primal_bound}

For a minimizer $\boldsymbol{\lambda}^{*}$ of a dual function $\delta$, let us denote by $\tilde{\mbf{R}}$ an optimal solution of the problem~\eqref{ch3:dual_fcn} obtained by applying Lemma~\ref{edm:lemma1}. Since $\sum_{k=1}^i \tilde{R}_k = f_{\beta}(\{1,2,\ldots,i\})$, and from Algorithm~\ref{alg:edm} and Theorem~\ref{thm:dilw}, $\sum_{k=1}^i R_k^{*}\leq, f_{\beta}(\{1,2,\ldots,i\})$,
it follows that
\begin{align}
\sum_{k=1}^{i-1} \tilde{R}_k-R_k^{*} \geq R_{i}^{*} - \tilde{R}_i. \label{edm:lemma:eq1}
\end{align}
By the formulation of the optimization problem~\eqref{ch3:primal},
the minimum value of the dual function $\delta$ is $R_i^{*}$. Therefore,
\begin{align}
\sum_{k=1}^{i-1} \lambda^{*}_k(\tilde{R}_k-R_k^{*}) = R_i^{*} - \tilde{R}_i. \label{edm:lemma:eq2}
\end{align}
From Algorithm~\ref{alg:edm} and Theorem~\ref{thm:dilw}, it follows that
\begin{align}
&~~~~\sum_{k=1}^i R_i^{*} \label{edm:lemma:eq3} \\
&= \min_{\mcl{P}}\left\{\sum_{\mcl{S}\in \mcl{P}}f_{\beta}(\mcl{S}) : \text{$\mcl{P}$ is a partition of}~\{1,2,\ldots,i\}\right\}. \nonumber
\end{align}
Let us denote by $\mcl{S}^{*}_i$, a set that belongs to an optimal partitioning $\mcl{P}^{*}$ w.r.t. problem~\eqref{edm:lemma:eq3} such that
$i \in \mcl{S}^{*}_i$. In this case, we have
\begin{align}
\sum_{k \in \mcl{S}^{*}_i} R_k^{*} = f_{\beta}(\mcl{S}_i^{*}). \label{edm:lemma:eq4}
\end{align}
Now, let us select $\boldsymbol{\lambda}^{*}$ as follows
\begin{align}
\lambda_k^{*}=
\begin{cases}
1 & \text{if}~k \in \mcl{S}_i^{*}, \label{edm:lemma:lambda} \\
0 & \text{otherwise.} 
\end{cases}
\end{align}
To verify that this choice of $\boldsymbol{\lambda}^{*}$ is indeed a dual optimal solution, note that from Lemma~\ref{edm:lemma1}, we have
\begin{align}
\sum_{k \in \mcl{S}_i^{*}} \tilde{R}_k = f_{\beta}(\mcl{S}_i^{*}).
\end{align}
Therefore,
\begin{align}
\sum_{k \in \mcl{S}_i^{*}} \tilde{R}_k - R_k^{*} = 0. \label{edm:lemma:eq5}
\end{align}
From~\eqref{edm:lemma:lambda} and~\eqref{edm:lemma:eq5}, it follows that 
\begin{align}
&\sum_{k \in \{1,\ldots i-1\} \setminus \mcl{S}_i^{*}} \lambda_k^{*}(\tilde{R}_k-R_k^{*}) \nonumber \\
&= R_i^{*}-\tilde{R}_i + \sum_{k \in \mcl{S}_i^{*} \setminus \{i\}}  \lambda_k^{*}(R_k^{*}-\tilde{R}_k). \label{edm:lemma:eq6}
\end{align}
This is consistent with~\eqref{edm:lemma:eq2}, and hence, $\boldsymbol{\lambda}^{*}$ is indeed a dual optimal solution. Therefore,
\begin{align}
\sum_{k=1}^{i-1} \lambda_k^{*} \leq i-1 \leq m.
\end{align}

\section{Proof of Lemma \ref{lm:cost_cap}}
\label{app:lm:cost_cap}
By Lemma~\ref{lm:f} we know that set functions $f_{N}$ and $f_{N+1}$, defined in~\eqref{fcn:f_star}, are fully submodular.
\begin{align}
f_{N}(\mcl{S})=
\begin{cases}
\text{rank}(\mbf{A}_{\mcl{S}}) & \text{if}~~\emptyset \neq \mcl{S} \subseteq \mcl{M}, \label{fcn:f_N} \\
0                                            & \text{if}~~\mcl{S}=\emptyset.
\end{cases}
\end{align}
\begin{align}
f_{N+1}(\mcl{S})=
\begin{cases}
1+\text{rank}(\mbf{A}_{\mcl{S}}) & \text{if}~~\emptyset \neq \mcl{S} \subseteq \mcl{M}, \label{fcn:f_N+1} \\
0                                            & \text{if}~~\mcl{S}=\emptyset.
\end{cases}
\end{align}
Let us denote by $\mbf{R}^{*}$ an optimal vector obtained by applying Algorithm~\ref{alg:allocation}.
From the correctness of Edmonds' algorithm, it directly follows that all faces of a submodular polyhedron $P(g_{\beta})$ are achievable, \emph{i.e.}, for any
$\mcl{S}\subseteq \mcl{M}$, there exists a rate vector $\mbf{R}$ such that $R(\mcl{S})=g_{\beta}(\mcl{S})$. Comparing $f_N$ and $f_{N+1}$, we see that all
``faces'' of polyhedron $P(f_{N+1})$ expended by $1$ compared to polyhedron $P(f_N)$ (and they are all achievable).
Hence, while applying Algorithm~\ref{alg:allocation} for $\beta=N+1$, we can see that the optimal rate vector $\tilde{\mbf{R}}$ will differ from $\mbf{R}^{*}$ in one coordinate. Let
\begin{align}
j^{*} = \argmin \left\{ d_i(R^{*}_i+1)~|~\mbf{R}^{*}+\mbf{e}(i) \in P(f_{N+1}) \right\}. \nonumber
\end{align}
Then,
\begin{align}
\tilde{R}_i=
\begin{cases}
R^{*}_i+1 & \text{if}~~i=j^{*}  \label{R:tilde}\\
R^{*}_i  & \text{otherwise}.
\end{cases}
\end{align}
Evaluating costs for $\beta=N$ and $\beta=N+1$, we obtain
\begin{align}
h(N)=\sum_{i=1}^m \varphi_i(R^{*}_i) = \sum_{i\neq j} \varphi_i(R^{*}_i) + \varphi_j(R^{*}_j). \label{fcn:h_N}
\end{align}
\begin{align}
h(N+1)=\sum_{i=1}^m \varphi_i(\tilde{R}_i) = \sum_{i\neq j} \varphi_i(R^{*}_i) + \varphi_j(R^{*}_j+1). \label{fcn:h_N_1}
\end{align}
Comparing~\eqref{fcn:f_N} and~\eqref{fcn:f_N+1}, we conclude that $h(N)\leq h(N+1)$ since $\varphi_j$ is a non-decreasing function.
Since $h$ is a convex function (see Theorem~\ref{thm:convex}), it immediately follows that \mbox{$\beta^{*}\leq N$}. 
\section{Proof of Theorem \ref{thm:transset}}\label{app:thm:transset}
Let us start by considering round $j=1$ of Algorithm~\ref{alg:allocation}. All rates are set to zero, \emph{i.e.}, $R^{*}_i=0$, $i=1,\ldots,m$.
To check whether user $i$ belongs to set $\mcl{T}_1$, we need to verify whether its update belongs to polyhedron $P(f_{\beta})$
\begin{align}
R^{*}(S)+1\leq f_{\beta}(\mcl{S}),~~\forall \mcl{S},~~\text{s.t.}~~i \in \mcl{S}, \label{app:exp1}
\end{align}
where $f_{\beta}$ is defined in~\eqref{fcn:f_star}. Since $\mbf{R}^{*}$ is a zero vector, we can write the condition~\eqref{app:exp1} as
\begin{align}
1\leq \beta-N+\text{rank}(\mbf{A}_{\mcl{S}}),~~\forall \mcl{S}\subseteq \mcl{M},~~\text{s.t.}~~i \in \mcl{S}, \label{app:exp2}
\end{align}
which is equivalent to
\begin{align}
1\leq \min_{i \in \mcl{S}\subseteq \mcl{M}} \{ \beta-N+\text{rank}(\mbf{A}_{\mcl{S}})\}. \label{app:exp3}
\end{align}
It is easy to see that $\mcl{S}=\{i\}$ is the minimizer of the above problem. Hence, $i\in \mcl{T}_1$ if
\begin{align}
\text{rank}(\mbf{A}_i) > N-\beta, \label{app:exp4}
\end{align}
which matches the theorem statement for $j=1$.

Say that user~$i$ belongs to $\mcl{T}_1$ and that he is scheduled to transmit in the first round according to the cost function.
Thus, user~$i$ transmits
\begin{align}
v_i^{(1)} = \mbf{u}^{(1)} \cdot \mbf{w},
\end{align}
where $\mbf{u}^{(1)}$ is appropriately chosen vector.
All the remaining users update their observation matrix by appending vector $\mbf{u}^{(1)}$  to it
\begin{align}
\mbf{A}_k \cup \mbf{u}^{(1)},~~\forall k\in \mcl{M}\setminus \{i\}. \label{app:exp5}
\end{align}

In the next round we reduce parameter $\beta$ by $1$, and again ask the same question whether user~$i$
belongs to $\mcl{T}_2$ for the updated set of observations. Combining~\eqref{app:exp4} and~\eqref{app:exp5} it is easy to see that in round $j$, the condition~\eqref{app:exp4} becomes
\begin{align}
\text{rank}\left(\mbf{A}_i \cup \mbf{u}^{(1)} \cup \cdots \cup \mbf{u}^{(j-1)} \right) > N-(\beta-j+1), \label{app:exp6}
\end{align}
which completes the proof.

\section{Proof of Theorem \ref{therem:restriction}}\label{app:theorem:restrictions}

Let $\mbf{R}$ be any feasible rate vector w.r.t. the problem~\eqref{de_sum4}, \emph{i.e.},
\begin{align}
R(\mcl{S})&\leq g_{\beta}(\mcl{S}),~~\forall \mcl{S}\subseteq \mcl{M}, \label{thm:restriction:eq1}  \\
R(\mcl{S})&\leq c(\mcl{S}),~~\forall \mcl{S}\subseteq \mcl{M}, \label{thm:restriction:eq2} \\
R(\mcl{M})&= g_{\beta}(\mcl{M})=\beta. \label{thm:restriction:eq8}
\end{align}
By substituting $\mcl{S}$ with $\mcl{M}\setminus \mcl{S}$ in~\eqref{thm:restriction:eq1}, we obtain
\begin{align}
R(\mcl{M}\setminus \mcl{S})&\leq g_{\beta}(\mcl{M} \setminus \mcl{S}),~~\forall \mcl{S}\subseteq \mcl{M}, \label{thm:restriction:eq7}
\end{align}
This can be rewritten as
\begin{align}
R(\mcl{S})&\geq R(\mcl{M}) - g_{\beta}(\mcl{M}\setminus \mcl{S}) \nonumber \\
          &=\beta - g_{\beta}(\mcl{M}\setminus \mcl{S}),~~\forall \mcl{S} \subseteq \mcl{M}, \label{thm:restriction:eq4}
\end{align}
where the last equality comes from ~\eqref{thm:restriction:eq8}
From~\eqref{thm:restriction:eq2}, \eqref{thm:restriction:eq8}, and~\eqref{thm:restriction:eq4} it follows that
\begin{align}
g_{\beta}(\mcl{M})-g_{\beta}(\mcl{M}\setminus \mcl{V}) \leq R(\mcl{V}) \leq c(\mcl{V}),~~\forall \mcl{V} \subseteq \mcl{M}. \label{thm:restriction:eq5}
\end{align}
From~\eqref{thm:restriction:eq5}, we have that
\begin{align}
g_{\beta}(\mcl{M})\leq g_{\beta}(\mcl{M}\setminus \mcl{V})+c(\mcl{V}),~~\forall \mcl{V} \subseteq \mcl{M}. \label{thm:restriction:eq6}
\end{align}
From \eqref{restriction} and~\eqref{thm:restriction:eq8}, we conclude that
\begin{align}
g_{\beta}^{\mbf{c}}(\mcl{M})=g_{\beta}(\mcl{M})=\beta. \label{thm:eq9}
\end{align}
Hence, $R(\mcl{M})=g_{\beta}^{\mbf{c}}(\mcl{M})$. Since $R_i\leq c_i$, it follows that
\begin{align}
R(\mcl{S})&=R(\mcl{V})+R(\mcl{S} \setminus \mcl{V}) \nonumber \\
&\leq g_{\beta}(\mcl{V})+c(\mcl{S} \setminus \mcl{V}),~\forall \mcl{V},\mcl{S}~\text{s.t.}~
\mcl{V}\subseteq \mcl{S} \subseteq \mcl{M}. \label{thm:eq7}
\end{align}
Finally~\eqref{thm:eq7} implies that
\begin{align}
R(\mcl{S})\leq \min \left\{ g_{\beta}(\mcl{V})+c(\mcl{S} \setminus \mcl{V})~|~\mcl{V}\subseteq \mcl{S} \right\},~\forall \mcl{S}\subseteq\mcl{M}. \nonumber
\end{align}
Hence, $\mbf{R}\in B(g_{\beta}^{\mbf{c}})$.

Conversely, let $\mbf{R}$ be such that  $\mbf{R} \in B(g_{\beta}^{\mbf{c}})$. Then,
\begin{align}
R(\mcl{S})&\leq g_{\beta}^{\mbf{c}}(\mcl{S}) \leq g_{\beta}(\mcl{S})+c(\emptyset)=g_{\beta}(\mcl{S}),~\forall \mcl{S}\subseteq \mcl{M}, \label{thm:eq4} \\
R(\mcl{S})&\leq g_{\beta}^{\mbf{c}}(\mcl{S}) \leq g_{\beta}(\emptyset)+c(\mcl{S})=c(\mcl{S}),~~\forall \mcl{S} \subseteq \mcl{M}, \label{thm:eq5} \\
R(\mcl{M})&= g_{\beta}^{\mbf{c}}(\mcl{M}) = \beta \label{thm:eq8}
\end{align}
where the second inequality in~\eqref{thm:eq4} and~\eqref{thm:eq5} directly follows from~\eqref{restriction}.
From~\eqref{thm:eq4},~\eqref{thm:eq5}, and \eqref{thm:eq8} it follows that
\begin{align}
\mbf{R} \in B(g_{\beta}),~~\text{s.t.}~~R_i\leq c_i,~~\forall i\in \mcl{M}.
\end{align}
This completes the proof. 

%\bibliographystyle{IEEEtran}
%\footnotesize{
%\bibliography{DatExc}

\end{document}